\documentclass[12pt,oneside,final]{amsart}

\usepackage[utf8x]{inputenc}

\usepackage{mystyle}
\DeclareMathAlphabet{\mathdutchcal}{U}{dutchcal}{m}{n}
\SetMathAlphabet{\mathdutchcal}{bold}{U}{dutchcal}{b}{n}
\DeclareMathAlphabet{\mathdutchbcal}{U}{dutchcal}{b}{n}
\newcommand{\ST}{\mathcal{S}}
\newcommand{\brstb}{\mathdutchcal{b}}
\newcommand{\brsts}{\mathdutchcal{s}}

\begin{document}

\title{Einstein-Hilbert gravity, higher derivatives and a scalar matter field \\}

\author{Klaus Sibold}

\address{Institute for Theoretical Physics, Leipzig University, Germany. 
\emph{Email address: }\rm \texttt{sibold(at)physik.uni-leipzig.de}.}

\date{\today}

\begin{abstract}
The present paper extends two previous one's on pure gravity dealing with 
Einstein-Hilbert and higher derivatives	by including a massless scalar
field as representative of matter. We study the
renormalization to all orders of perturbation theory, provide the 
Slavnov-Taylor identity, symmetric partial differential equations
and derive finiteness properties in the Landau gauge. It is shown that
beginning with one-loop negative norm states originating from higher 
derivatives disappear.

\end{abstract}

\maketitle

\tableofcontents
 
\section{Introduction}
In two previous papers Steffen Pottel and the present author discussed
the quantization of Einstein-Hilbert gravity (EH) in a perturbative framework
[PS21,PS23].
The existence of Green's functions was ensured by adding higher derivative terms
(hds) to the action.
Those introduce negative norm into the state space and require quite some
effort to arrive at meaningful quantities. Although those problems have not yet
been fully solved it seems reasonable to study in addition the effect
of matter: gravity and matter have to coexist. Here we add a scalar field
as representative of matter. \\
For the benefit of the reader we shall take over at appropriate places
not only results of the previous papers, but also main parts of the derivations.
Hopefully we arrive in this way at a reasonable view of the entire subject.\\

For the scalar field we assume that it is massless and invariant under rigid 
Weyl transformation. The reason is simple: for such a field Elisabeth Kraus
and the present author studied the behaviour of a specifically defined
energy-momentum tensor under all transformations of the conformal group to 
all orders of perturbation theory \cite{EKKSI}. The coupling to an external symmetric tensor
field $h^{\mu\nu}$ was as crucial as the gauging of translations. It
turned out that $g^{\mu\nu}= \eta^{\mu\nu}+h^{\mu\nu}$ and its covariant
companion $g_{\mu\nu}$ were the only solutions of the Ward identity of
local translations up to field redefinitions of $h^{\mu\nu}$ as a function
of itself and had to transform as gravity requires it for a metric. Hence one 
arrived just by Noether's procedure at gravity \cite{EKKSII,EKKSIII}. Permitting
thereafter propagation of this gravity field yields by sheer power counting
as theory to be studied not only the Einstein-Hilbert term $\int\sqrt{-g}R$,
but also $\int\sqrt{-g}R^2$ and 
$\int\sqrt{-g}R^{\mu\nu}R_{\mu\nu}$. Such
a model is power counting renormalizable and closes in itself under
renormalization. The most obvious understanding for this claim is provided
by the field equation for $g^{\mu\nu}$: the energy-momentum tensor of the
scalar field which has canonical dimension four appears on an equal footing 
with the pure gravity terms. Hence all of the mentioned one's have to be
admitted if one thinks in terms of renormalization.\\

In the present paper we study the renomalization of this model to all orders of
perturbation theory. For this we provide first power counting and convergence of 
our scheme. Then we establish the Slavnov-Taylor (ST) identity, derive the 
symmetric partial differential equations and
prove noteworthy finiteness properties in the Landau gauge. The Ward identity 
for rigid Weyl transformations is discussed and seen to replace the
Callan-Symanzik equation of flat spacetime. Important for interpretation and
understanding of the model is the result that beginning with
one-loop, the second poles in the propagator $<hh>$ are absent.
This implies that beginning with one-loop negative norm contributions which
had been introduced by the higher derivative terms disappear. Those survive
however at tree level. In the conclusions we speculate on how they might
be possibly removed.\\

\section{Tree approximation}
%file TreeApproximation.tex\\
For a decent perturbative treatment it is mandatory to set up the first orders
carefully. In the present context this refers to the zero-loop order and the
first and second order in the number of fields.

\subsection{The model and its invariances}
As in the study of pure EH we have also in the present case with an
additional scalar matter field to include invariants under diffeomorphisms
up to fourth order in the derivatives in order to achieve power counting 
renormalizability. Since we restrict our considerations to spacetimes which are 
topologically equivalent to flat one's, it suffices to start with 
\begin{eqnarray}\label{ivc}
	\Gamma^{\rm class}_{\rm inv}&=&\dm\Bigl(
	c_3\kappa^{-2}R+c_2R^2+c_1R^{\mu\nu}R_{\mu\nu}\Bigr)\\
	&&\quad\quad+\int\Bigl((-g)^{1/4}\frac{c_R}{2}\varphi^2R
+(-g)^{1/4}\frac{1}{2}g^{\mu\nu}\mathcal{D}_\mu\varphi\mathcal{D}_\nu\varphi
	-\frac{\lambda}{4!}\varphi^4\Bigr), 
\end{eqnarray}
i.e.\ we can omit the cosmological term. Its absence in all orders will be 
guaranteed by a normalizazion condition. $\kappa$ denotes the gravitational 
constant, $R_{\mu\nu},R$ the Ricci 
curvature tensor, resp.\ scalar. The second line comprises the invariants 
of dimension four which can be built with a massless scalar field of canonical 
dimension one. Their peculiar weight factors occur because we have given 
$\varphi$ the $\brsts$ transformation law (\ref{vrps}). That in turn we introduced 
in \cite{EKKSI} for local translations such that all transformations under the 
conformal group can be obtained via $x$-moments of local translations.
This invariance under local translations, which is nothing but 
general coordinate transformations, is to be translated into
Becchi-Rouet-Stora-Tyutin invariance (BRST) with respective gauge fixing.
The field $h^{\mu\nu}$ is defined via
\be\label{dfh}
h^{\mu\nu}=g^{\mu\nu}-\eta^{\mu\nu} .
\ee
The propagators of $h$ (s.b.) will tell us that 
$h$ has canonical dimension $0$, hence $\kappa$ must not show up in its definition.

The classical action 
\begin{eqnarray}\label{clssct}
\Gamma^{\rm class}&=& \Gamma^{\rm class}_{\rm inv} + \Gamma_{\rm gf} 
				  + \Gamma_{\phi\pi}+\Gamma_{\rm e.f.}\\
\Gamma_{\rm gf}&=&-\frac{1}{2\kappa}\int g^{\mu\nu}
                                (\partial_\mu b_\nu+\partial_\nu b_\mu) 
	                     -\frac{1}{2}\alpha_0\int \eta^{\mu\nu}b_\mu b_\nu \label{gf1} \\
	\Gamma_{\phi\pi}&=&-\frac{1}{2}\int(D^{\mu\nu}_\rho c^\rho)
      (\partial_\mu \bar{c}_\nu +\partial_\nu\bar{c}_\mu)\label{gf2}\\
D^{\mu\nu}_\rho&\equiv&-g^{\mu\lambda}\delta^\nu_\rho\partial_\lambda
		       -g^{\nu\lambda}\delta^\mu_\rho\partial_\lambda
		       +\partial_\rho g^{\mu\nu}\\
	\Gamma_{\rm e.f.}&=&\int (K_{\mu\nu}\brsts h^{\mu\nu}+L_\rho 
	\brsts c^\rho	                    +Y \brsts \varphi)
\end{eqnarray}
is invariant under the BRST-transformation 
\begin{eqnarray}\label{brst}
	\brsts g^{\mu\nu}&=&\kappa D^{\mu\nu}_\rho c^\rho
	\qquad\quad \brsts c^\rho=-\kappa c^\lambda \partial_\lambda c^\rho	 \qquad 
	\brsts \bar{c}_\rho= b_\rho
	\quad \brsts b_\rho=0\\
	\brsts_0 h^{\mu\nu}&=&-\kappa(\partial^\mu c^\nu+\partial^\nu c^\mu)
	\quad \brsts_1 h^{\mu\nu}=
	      -\kappa(\partial_\lambda c^\mu h^{\lambda\nu}
			     +\partial_\lambda c^\nu h^{\lambda\mu}
			     -c^\lambda\partial_\lambda h^{\mu\nu})\\
	\brsts\varphi&=&c^\lambda\partial_\lambda\varphi
	      +\frac{1}{4}\partial_\lambda c^\lambda\varphi \label{vrps}
\end{eqnarray}
In accordance with the expansion in the number of fields we
have introduced the transformations $\brsts_0,\brsts_1$ which maintain the number, resp.\ raise it by one.
$K_{\mu\nu}, L_{\rho},Y$ are external fields to be used for generating insertions 
of non-linear field transformations. The Lagrange multiplier 
$b_\mu$ 
couples to $\partial_\lambda h^{\mu\lambda}$ and thus
fixes eventually these derivatives (deDonder like gauge fixing).

\subsection{Propagators}
For later convenience we quote here from paper I the bilinear terms of 
$\Gamma^{\rm class}_{\rm inv}$ and add that of the scalar field, all in Fourier 
space. 

\begin{eqnarray}\label{bln}
	\Gamma_{h_{\mu\nu}h_{\rho\sigma}}&=& 
\frac{1}{4}\sum_{KLr}\gamma^{(r)}_{KL}(p^2)(P_{KL}^{(r)})_{\mu\nu\rho\sigma} \\
	\Gamma_{b_\rho h_{\mu\nu}}&=&
-\frac{i}{\kappa} \Big(\frac{1}{2}(\theta_{\rho\mu} p_\nu+\theta_{\rho\nu}p_\mu) 
                                                 +\omega_{\mu\nu}p_\rho \Big) \\
	\Gamma_{b_\rho b_\sigma}&=& -\alpha_0\eta_{\rho\sigma} \\
	\Gamma_{c_\rho\bar{c}_\sigma}&=& -ip^2 \big( \theta_{\rho\sigma}\xi(p^2)+
	\omega_{\rho\sigma}\frac{1}{2}\eta(p^2) \big) \\
	\Gamma_{\varphi\varphi} &=&-p^2.
\end{eqnarray}
For the $h$-bilinear terms we introduce projection operators $P$ (see App.\ A)
and general coefficient functions $\gamma$. Their values read in tree 
approximation
\begin{eqnarray}\label{coffs}
\gamma^{(2)}_{TT} &=&-p^2(c_1p^2-c_3\kappa^{-2})\\
\gamma^{(0)}_{TT} &=&p^2 \big((3c_2+c_1)p^2-2c_3\kappa^{-2} \big)\\ 
\gamma^{(1)}_{SS}&=&\gamma^{(0)}_{WW}=\gamma^{(0)}_{TW}=\gamma^{(0)}_{WT}=0 .
\end{eqnarray}
The coefficients of $\Gamma_{bh}$ and
$\Gamma_{bb}$ turn out to be fixed, whereas those of $\Gamma_{c\bar{c}}$
can be very general with tree values $\xi=\eta=1$.\\

For the $\langle hh \rangle$-propagators we introduce like for the 2-point-vertex 
functions an expansion in terms of projection operators
\be\label{hhpropproj}
	G^{hh}_{\mu\nu\rho\sigma}= 
		4\sum_{KLr} \langle hh \rangle^{(r)}_{KL}(P_{KL}^{(r)})_{\mu\nu\rho\sigma}.
\ee
The gauge parameter independent solutions $\langle hh \rangle^{(r)}_{KL}$ turn out to be 
\be\label{bosprop}
	\langle hh \rangle^{(2)}_{TT}=\frac{i}{\gamma^{(2)}_{TT}}			       
	\qquad\quad\langle hh \rangle^{(0)}_{TT}=\frac{i}{\gamma^{(0)}_{TT}},		       
\ee
whereas the ``gauge parameter multiplet'' is given by
\begin{eqnarray}\label{gaugeprop}
	\langle hh \rangle^{(1)}_{SS}&=&\frac{4i\alpha_0\kappa^2}{p^2} \qquad
	\langle hh \rangle^{(0)}_{WW}=\frac{4i\alpha_0\kappa^2}{p^2}\\		       
  \qquad \langle hh \rangle^{(0)}_{TW}&=&\langle hh \rangle^{(0)}_{WT}=0.       
\end{eqnarray}
The gauge parameter independent part is determined by the coefficient
functions $\gamma$, which depend on the model, i.e.\ by the invariants
and by higher orders, whereas the gauge multiplet is essentially fixed and 
determined by the specific gauge choice. The remaining bosonic propagators read
\be\label{bprop}
\langle b_\rho h_{\mu\nu} \rangle =\frac{\kappa}{p^2} \big( (p_\mu \theta_{\nu\rho}+p_\nu 
      \theta_{\mu\rho})b_1+p_\rho \omega_{\mu\nu}b_2+p_\rho \theta_{\mu\nu}b_3 \big)
\quad \mbox{and}\quad \langle b_\rho b_\sigma\rangle =0   .   
\ee
In the tree approximation $b_1=b_2=1$ and $b_3=0$.
The antighost/ghost propagator has the general form 
\be\label{ggpropKO}
\langle \bar{c}_\rho c_\sigma \rangle=\frac{-1}{p^2}
    \Big( \frac{\theta_{\rho\sigma}}{\xi(p^2)}+\frac{\omega_{\rho\sigma}}{\eta(p^2)} \Big).
\ee
The tree approximation values are $\xi=\eta=1$, s.t.\
\be\label{ggpropEK}
\langle \bar{c}_\rho c_\sigma \rangle =-i \big(\theta_{\rho\sigma}
                        +\frac{1}{2}\omega_{\rho\sigma} \big) \frac{1}{p^2}.
\ee
The propagator of the scalar field is simply   
\be\label{scpr}
\langle \varphi\varphi\rangle= \frac{i}{p^2}
\ee

\subsection{The Slavnov-Taylor identity in tree approximation}
Since the $\brsts$-variations of $h,c,\varphi$ are non-linear in the fields, they are best implemented in higher orders via coupling to external fields 
(cf. \eqref{clssct}), hence the ST identity then reads 
\be\label{fbrst}
\mathcal{S}(\Gamma)\equiv
\int(\frac{\delta\Gamma}{\delta{K}}\frac{\delta\Gamma}{\delta h}
+\frac{\delta\Gamma}{\delta L}\frac{\delta\Gamma}{\delta c}
+\frac{\delta\Gamma}{\delta Y}\frac{\delta\Gamma}{\delta\varphi}
+b\frac{\delta\Gamma}{\delta\bar{c}})=0 .
\ee
Since the $b$-equation of motion
\be\label{beq}
\frac{\delta \Gamma}{\delta b^\rho}=
                     \kappa^{-1}\partial^\mu h_{\mu\rho}-\alpha_0b_\rho
\ee
is linear in the quantized fields, it can be integrated trivially to the 
original
gauge fixing term. Thus it turns out to be useful to introduce a functional
$\bar{\Gamma}$ which does no longer depend on the $b$-field:
\be\label{Gmmbr}
\Gamma=\Gamma_{\mathrm{gf}}+\bar{\Gamma} .
\ee
One finds
\be\label{rstc}
\kappa^{-1}\partial_\lambda\frac{\delta\bar{\Gamma}}{\delta K_{\mu\lambda}}
+\frac{\delta\bar{\Gamma}}{\delta\bar{c}_\mu} =0
\ee
as restriction. Hence $\bar{\Gamma}$ depends on $\bar{c}$ only via
\be\label{sceH}
H_{\mu\nu}=K_{\mu\nu} - \frac{1}{2\kappa}(\partial_\mu\bar{c}_\nu+\partial_\nu\bar{c}_\mu)
\ee
and the ST identity takes the form  
\begin{eqnarray}\label{brGm}
\mathcal{S}(\Gamma)&=&\frac{1}{2}\mathcal{B}_{\bar{\Gamma}}\bar{\Gamma}=0\\
	\mathcal{B}_{\bar{\Gamma}}&\equiv& 
	\int(
  \frac{\delta\bar{\Gamma}}{\delta H}\frac{\delta}{\delta h}
+ \frac{\delta\bar{\Gamma}}{\delta h}\frac{\delta}{\delta H} 
+ \frac{\delta\bar{\Gamma}}{\delta L}\frac{\delta}{\delta c}
+ \frac{\delta\bar{\Gamma}}{\delta c}\frac{\delta}{\delta L}
+ \frac{\delta\bar{\Gamma}}{\delta Y}\frac{\delta}{\delta \varphi}
+ \frac{\delta\bar{\Gamma}}{\delta \varphi}\frac{\delta}{\delta Y}) .
\end{eqnarray}
This form shows that $\mathcal{B}_{\bar{\Gamma}}$ can be interpreted as a variation und thus
(\ref{brGm}) expresses an invariance for $\bar{\Gamma}$.

\subsection{Unitarity in the tree aproximation}
The $S$-operator can be defined via
\begin{eqnarray}\label{sma}
S&=&:\Sigma: Z(\uJ)|_{\uJ=0}
	\qquad \Sigma\equiv\exp\left\{ {\int dx\,dy\, \Phi_{\rm in}(x)K(x-y)z^{-1}
	\frac{\delta}{\delta \uJ(y)}}\right\}, 
\end{eqnarray}
where $\uJ$ denotes the sources 
$J_{\mu\nu},j_{\bar{c}}^\rho,j_c^\rho,j^\rho_b,j_\varphi$ for the fields 
$h^{\mu\nu},\bar{c}^\rho,c^\rho,b^\rho,\varphi$, respectively, and
their in-field versions are collected in $\Phi_{\rm in}.$ 
$K(x-y)z^{-1}$ refers to all in-fields and stands for the higher derivative
wave operator, hence removes the complete (tree approximation) propagator matrix.
$:\Sigma:$ would then map onto the respective large Fock space of the higher 
derivative model. In order to clarify the structure of the statespace
and to identify first the one of pure EH, we put $c_1=c_2=0$.
Now we study the unphysical degrees of freedom which go along with that 
model. They differ slightly from those studied by \cite{KuOj} because we employ 
a different field $h$, but the general structure is the same 
(cf.\ eqs. (353)..(356) in \cite{pottel2021perturbative}).
Here we follow %\cite{LHB}           
\cite{Becchi:1985bd}
and would like to show, that the $S$-matrix
commutes with the BRST-charge $Q$ by establishing the equations 
\be\label{brscomm}
[\mathcal{S},:\Sigma:]Z_{|\uJ=0}=-[Q,:\Sigma:]Z_{|\uJ=0}=[Q,S]=0,
\ee
where  
\begin{equation}
    \mathcal{S}\equiv \int \Big(J_{\mu\nu}\frac{\delta}K_{\mu\nu}
                           -j^\rho_c\frac{\delta}{\delta L^\rho}
                           -j_\varphi\frac{\delta}{\delta Y}
 	  -j^\rho_{\bar{c}}\frac{\delta}{\delta j^\rho_b} \Big)
 	  \quad \mbox{with} \quad
 	  \mathcal{S}Z=0 \, .
\end{equation}
The lhs of \eqref{brscomm} is a commutator in the space of functionals, i.e.\ of $\mathcal{S}$, the ST-operator, with the $S$-matrix defined on the functional level via $Z$, the generating functional for general Green functions. Now
\be\label{2brscomm}
[\mathcal{S},:\Sigma:]Z_{|\uJ=0}=0 
\ee
since the first term of the commutator vanishes because $\mathcal{S}=0$ for
vanishing sources, the second term of the commutator vanishes due to the
validity of the ST-identity.\\
The rhs of (\ref{brscomm}) is an equation in terms of (pre-)Hilbert 
space operators: $S$-operator and BRST-charge, both defined on the indefinite
metric Fock space of creation and annihilation operators. The claim is 
that we can find an operator $Q$ such that the rhs holds true.\\
We then know that a subspace defined by $Q|\mathrm{phys}\rangle=0$ is stable under $S$, hence
physical states are mapped into physical states.\\
To show that (\ref{2brscomm}) indeed holds, we observe first that the commutator
$[\mathcal{S},:\Sigma:]$ is of the form $[\mathcal{S},e^X]$. If 
$[\mathcal{S},X]$ commutes with $X$, one can reorder
the series into $[\mathcal{S},e^X]=[\mathcal{S},X]e^X$. This has to be 
evaluated.
Since in the tree approximation $z=1$, hence 
$K(x-y)_{\Phi\Phi'}= \Gamma_{\Phi\Phi'}$
we define for the explicit calculation  
\begin{eqnarray}\label{auxy}
X\equiv&&\int\Big(
h^{\mu\nu}\Gamma^{hh}_{\mu\nu\rho\sigma}\frac{\delta}{\delta J_{\rho\sigma}}
+h^{\mu\nu}\Gamma^{hb}_{\mu\nu\rho}\frac{\delta}{\delta j_\rho^b}
+\varphi\Gamma^{\varphi\varphi}\frac{\delta}{\delta j_\varphi}   \\
&&\quad +\, b^{\rho}\Gamma^{bh}_{\rho\alpha\beta}\frac{\delta}{\delta J_{\alpha\beta}}
+ b^{\rho}\Gamma^{bb}_{\rho\sigma}\frac{\delta}{\delta j_\sigma^b}
+ c^{\rho}\Gamma^{c\bar{c}}_{\rho\sigma}\frac{\delta}{\delta j_\sigma^{\bar{c}}}
+ \bar{c}^{\rho}\Gamma^{\bar{c}c}_{\rho\sigma}\frac{\delta}{\delta j_\sigma^c}\Big) .
\end{eqnarray}
For the desired commutator one finds 
\be\label{XYcomm}
[\mathcal{S},X]=-\int\Big(
h^{\mu\nu}\Gamma^{hh}_{\mu\nu\rho\sigma}\frac{\delta}{\delta K_{\rho\sigma}}
+\varphi\Gamma^{\varphi\varphi}\frac{\delta}{\delta Y}
-c^\rho\Gamma^{c\bar{c}}_{\rho\sigma}\frac{\delta}{\delta j^b_\sigma}
-\bar{c}^\rho\Gamma^{\bar{c}c}_{\rho\sigma}\frac{\delta}{\delta L_\sigma}\Big),
\ee
so it clearly commutes with $X$.\\
In the next step we have to consider $:[\mathcal{S},X]e^X:Z$, i.e.\ terms of 
the type  
\begin{align}\label{XYcomm1}
-\int:\Big(h^{\mu\nu}\Gamma^{hh}_{\mu\nu\rho\sigma}
	\frac{\delta}{\delta K_{\rho\sigma}}
	+\varphi\Gamma^{\varphi\varphi}\frac{\delta}{\delta Y}
-c^\rho\Gamma^{c\bar{c}}_{\rho\sigma}\frac{\delta}{\delta j^b_\sigma}
	-\bar{c}^\rho\Gamma^{\bar{c}c}_{\rho\sigma}\frac{\delta}{\delta L_\sigma} \Big)&:
	X(1)\cdots X(n)\cdot Z(\uJ)_{|\uJ=0} \\
\intertext{ i.e.}
-\int:\Big(h^{\mu\nu}\Gamma^{hh}_{\mu\nu\rho\sigma}\kappa D^{\rho\sigma}_\lambda 
	                                           c^\lambda
+\varphi\Gamma ^{\varphi\varphi}(c^\lambda\partial_\lambda\varphi
	+\frac{1}{4}\partial_\lambda c^\lambda\varphi)
	-c^\rho\Gamma^{c\bar{c}}_{\rho\sigma}b^\sigma&\\ \nonumber
-\bar{c}^\rho\Gamma^{\bar{c}c}_{\rho\sigma}c^\lambda\partial_\lambda c^\sigma\Big):
	X(1)\cdots X(n)\cdot Z(\uJ)_{|\uJ=0} . & 
\end{align}	
These terms constitute insertions into the functional $Z$. A closer look 
in terms of Feynman diagrams reveals that due to momentum conservation
from $D^{\rho\sigma}_\lambda c^\lambda$ only terms linear in the fields
survive, neither the matter nor the last term bilinear in $c$ contribute -- when
going on mass shell they cannot develop particle poles. We arrive thus at
\be\label{transeq}
	:[\mathcal{S},X]:Z=
     :\Sigma\Big[ \int(-h^{\mu\nu}\Gamma^{hh}_{\mu\nu\alpha\beta}
                 \kappa (\partial^\alpha c^\beta+\partial^\beta c^\alpha)
	 +c^\rho\Gamma^{c\bar{c}}_{\rho\sigma}b^\sigma)
       \Big]:\cdot Z(\underline{J})_{|\underline{J}=0}.
\ee
The second factors in the insertion are just the linearized BRST-variations
of $h^{\alpha\beta}$, resp.\ $\bar{c}^\sigma$. This suggests to introduce
 a corresponding BRST operator $Q$ which generates these
transformations
\begin{eqnarray}\label{linBRS}
	Q\Gamma&\equiv&\int\Big\lbrack
	\kappa(\partial^\mu c^\nu+\partial^\nu c^\mu)
	\frac{\delta}{\delta h^{\mu\nu}}
		   + b^\rho\frac{\delta}{\delta \bar{c}^\rho}\Big\rbrack \Gamma\\
	QZ_c&\equiv&-\int\Big\lbrack
	\kappa(\partial^\mu\frac{\delta Z_c}{\delta j_\nu^c} +	
	\partial^\nu\frac{\delta Z_c}{\delta j_\mu^c})J_{\mu\nu}+
	\frac{\delta Z_c}{\delta j^b_\rho}j_\rho^{\bar{c}}
	\Big\rbrack\\
	QZ&\equiv&-i\int\Big\lbrack
	J_{\mu\nu}\kappa(\partial^\mu\frac{\delta}{\delta j^c_\nu} +	
		   \partial^\nu\frac{\delta}{\delta j_\mu^c})+
	j^{\bar{c}}_\rho\frac{\delta}{\delta j^b_\rho}\Big\rbrack Z ,
\end{eqnarray}
and to calculate the commutator $[Q,:\Sigma:]Z_{|\uJ=0}$. And, indeed it
coincides with the rhs of (\ref{transeq}). Following in detail the 
aforementioned diagrammatic analysis we have a simple interpretation: 
in the Green functions $G(y;z_1,...,z_n)$ a field
entry has been replaced by the linearized BRST-transformation of it.
Having established (\ref{brscomm}) one can continue along the lines of 
\cite{KuOj}, form within the linear subspace of physical states equivalence
classes by modding out states with vanishing norm with the well-known result that
these factor states have non-vanishing norm and the $S$-matrix is unitary.\\
It would now be most natural to extend these considerations by taking into
account the additional degrees of freedom going along with the higher
derivatives. This requires however identifying those and understanding their
behaviour in higher orders. This will be done below, Sect.\ 10.\\

\subsection{Parametrization and gauge parameter independence.}
It is a necessary preparation for higher orders to clarify, which parameters the model contains and how they are fixed.
Also a glance at the free propagators, (\ref{bosprop}) versus
(\ref{gaugeprop}), shows that they differ in their fall-off properties depending from the value of the gauge parameter 
$\alpha_0$. Since Landau gauge $\alpha_0=0$ simplifies calculations enormously we would like to show that it is stable against perturbations. Since these two issues are closely linked
we treat them here together.
Obvious parameters are the couplings $c_0,c_1,c_2,c_3,c_{\rm R},\lambda$. In 
the next subsection we give a prescription, how to fix them by
appropriate normalization conditions. Also obvious is the gauge parameter $\alpha_0$. It will be fixed by the equation of motion for the $b$-field. Since this equation is linear in the $b$-field it also determines its amplitude.
Less obvious is the normalization of the fields
$h^{\mu\nu},c^\rho$ and of the external fields $K,L,Y$. In order to fix
their amplitudes it is convenient to inquire under which linear redefinitions 
of them the ST (\ref{fbrst}) stays invariant.  
We define  
\begin{align}\label{frdfs}  
\hat{h}^{\mu\nu}&=z_1(\alpha_0)h^{\mu\nu} \qquad \hat{\varphi}=t_\varphi(\alpha_0) 
                                           &\hat{c}^\rho&=y(\alpha_0)c^\rho\\
\hat{K}_{\mu\nu}&=\frac{1}{z_1(\alpha_0)}K_{\mu\nu} 
                                  \quad \hat{Y}=\frac{1}{t_\varphi(\alpha_0)}  
                                    &\hat{L}_\rho&=\frac{1}{y(\alpha_0)}L_\rho, 
\end{align}
where we admitted a dependence on the gauge parameter because we
would like to vary it and detect in this way $\alpha_0$-dependence algebraically. 
Clearly, the values for $z_1,t_\varphi$ and $y$ have to be prescribed. It is also 
clear that with 
$\alpha_0$-independent values for $z_1,t\varphi$ and $y$ the ST-identity is 
maintained.
In order to make changes of $\alpha_0$ visible we differentiate (\ref{clssct}) with respect to it, i.e.\  
\be\label{varalph}
\frac{\partial}{\partial\alpha_0}\Gamma=\frac{\partial}{\partial{\alpha_0}}
        \Gamma_{\rm gf}=\int(-\frac{1}{2})b_\mu b_\nu\eta^{\mu\nu}=
	\brsts\int(-\frac{1}{4})(\bar{c}_\mu b_\nu+\bar{c}_\nu b_\mu)\eta^{\mu\nu} .
\ee
We observe that this is an $\brsts$-variation and thus, if we introduce a fermionic partner $\chi= \brsts\alpha_0$ and perform the change
\be\label{chG}
\Gamma_{\rm gf} +\Gamma_{\phi\pi}\to\Gamma_{\rm gf}+\Gamma_{\phi\pi}
    +\int(-\frac{1}{4})\chi(\bar{c}_\mu b_\nu+\bar{c}_\nu b_\mu)\eta^{\mu\nu} .
\ee
we have
\be\label{chidouble}
\ST(\Gamma)+\chi\partial_{\alpha_0}\Gamma=0 .
\ee
We carry over this extended BRST-transformation to $Z$
\be\label{chiZ}
\hat{\mathcal{S}}Z\equiv\mathcal{S}Z+\chi\partial_{\alpha_0}Z=0,
\ee
with the implication
\be
\partial_\chi(\hat{\mathcal{S}}Z)=0 \quad\Rightarrow\quad 
    \partial_{\alpha_o} Z=-\mathcal{S}\partial_\chi Z 
\ee
showing that $\alpha_0$-dependence is a BRST-variation, hence unphysical. 
This last equation
can be easily checked on the free propagators (for propagators connected and 
general Green functions coincide).

Using for $Z(\uJ)$ the form
\be\label{znt1}
Z(\uJ)=\exp \Big\{i\int \mathcal{L}_{\rm int}\big(\frac{\delta}{i\delta\uJ}\big)\Big\}Z_0 \qquad Z_0=\exp\Big\{\int i\uJ \langle \Phi\Phi \rangle i\uJ \Big\}
\ee
one obtains
\be\label{znt2}
\partial_{\alpha_0}Z(\uJ)=\partial_{\alpha_0}Z_0\cdot Z(\uJ)
=\Big(\partial_{\alpha_0}\int i\uJ \langle \Phi\Phi \rangle i\uJ \Big)\cdot Z .
\ee
(Here $\uJ$ stands for the sources of all propagating fields 
$\Phi$.)
Hence $\alpha_0$-dependence remains purely at external lines, if one does not add
$\alpha_0$-dependent counterterms, and then vanishes on the $S$-matrix where these
lines are amputated. It also means that the power counting for the gauge multiplet
is irrelevant because this multiplet shows up only as external 
lines.

We now step back and analyse $\alpha_0$-dependence more systematically. Equations (\ref{chidouble}), (\ref{chiZ}) and the analogous one for connected Green functions
\be\label{chiZc}
\mathcal{S}Z_c+\chi\partial_{\alpha_0}Z_c=0,
\ee
where $\alpha_0$ undergoes the change
\be\label{vrgp}
\brsts\alpha_0=\chi \qquad \brsts\chi=0
\ee
have to be solved. 
The rhs of (\ref{chG}) is solution of the extended gauge condition
\be\label{egc}
\frac{\delta \Gamma}{\delta b^\rho}=
\kappa^{-1}\partial^\mu h_{\mu\rho} -\alpha_0 b^\nu-\frac{1}{2}\chi\bar{c}_\rho.
\ee 
Acting with $\delta/\delta b^\rho$ on the ST (\ref{chidouble}) we find
that the ghost equation of motion has changed accordingly
\be\label{gee}
G^\rho\Gamma\equiv \Big(\kappa^{-1}\partial^\mu\frac{\delta}{\delta K_{\mu\rho}}
  +\frac{\delta}{\delta \bar{c}_\rho} \Big) \Gamma=\frac{1}{2}\chi b^\rho
\ee
As in (\ref{Gmmbr}) and (\ref{sceH}) we introduce 
$H_{\mu\nu}=K_{\mu\nu} - \frac{1}{2\kappa}(\partial_\mu\bar{c}_\nu
     +\partial_\nu\bar{c}_\mu)$ and $\bar{\Gamma}$ by
\be\label{pbrGm}
\Gamma=\bar{\Gamma}
      +\int \Big(-\frac{1}{2}\alpha_0 b_\mu b_\nu \eta^{\mu\nu}
  -\frac{1}{2\kappa}h^{\mu\nu}(\partial_\mu b_\nu+\partial_\nu b_\nu)
-\frac{1}{4}\chi(\bar{c}_\mu b_\nu+\bar{c}_\nu b_\mu)\eta^{\mu\nu} \Big) .
\ee
The extended ST reads in terms of $\bar{\Gamma}$
\be\label{ebrG}
\ST (\Gamma)=\mathcal{B}(\bar{\Gamma})=0
\ee
with  
\be\label{pbrGm2}
\mathcal{B}(\bar{\Gamma})\equiv
\int \Big(\frac{\delta\bar{\Gamma}}{\delta K}\frac{\delta\bar{\Gamma}}
                                                      {\delta h}
      +\frac{\delta\bar{\Gamma}}{\delta Y}\frac{\delta\bar{\Gamma}}
                                                      {\delta\varphi}     
      +\frac{\delta\bar{\Gamma}}{\delta L}\frac{\delta\bar{\Gamma}}
                                                      {\delta c}
      +\chi\frac{\partial\bar{\Gamma}}{\partial \alpha_0} \Big).
\ee
$\bar{\Gamma}$ satisfies the homogeneous ghost equation of motion
\be\label{pgem}
G\bar{\Gamma}=0.
\ee
We now have to find the most general solution of ghost equation
(\ref{gee}) and the new ST (\ref{ebrG}). Due to dimension and
$\phi\pi$-charge neutrality $\bar{\Gamma}$ can be decomposed
as  
\be\label{gsa}
\bar{\Gamma}=\bar{\bar{\Gamma}}(h,c,K,L,Y,\alpha_0)
                 +\chi\int(f_K(\alpha_0)Kh+f_Y(\alpha_0)Y\varphi+f_L(\alpha_0)Lc)
\ee            
With the choice of linear dependence from $h$, however, we
certainly do not cover the most general case:     
due to the vanishing dimension of $h^{\mu\nu}$ one could 
replace the linear factor $h^{\mu\nu}$ by an arbitrary function 
$\mathcal{F}^{\mu\nu}(h)$ in $K_{\mu\nu}h^{\mu\nu}$.
For simplicity we discuss here the linear case, which continues (\ref{frdfs}), 
whereas for the non-linear we refer to the discussion in [PS21], Section VIII.\\

From (\ref{ebrG}) and (\ref{pbrGm2}) we deduce that  
\begin{eqnarray}\label{ntrm}
0=&&\mathcal{B}(\bar{\Gamma})=\mathcal{B}(\bar{\bar{\Gamma}})|_{\chi=0}+\chi\int(
-f_Hh^{\mu\nu}\frac{\delta\bar{\bar{\Gamma}}}{\delta h^{\mu\nu}}
+f_HH^{\mu\nu}\frac{\delta\bar{\bar{\Gamma}}}{\delta H^{\mu\nu}}\\
&&-f_Y\varphi\frac{\delta\bar{\bar{\Gamma}}}{\delta\varphi}
+f_YY\frac{\delta\bar{\bar{\Gamma}}}{\delta Y}
+f_Lc\frac{\delta\bar{\bar{\Gamma}}}{\delta c}
-f_LL\frac{\delta\bar{\bar{\Gamma}}}{\delta L})
+\chi\frac{\partial\bar{\bar{\Gamma}}}{\partial \alpha_0} .
\end{eqnarray}
At $\chi=0$ follows first
\be\label{ntrm2}
\mathcal{B}(\bar{\bar{\Gamma}})|_{\chi=0}=0,
\ee
and then  
\begin{eqnarray}\label{ntrm3}
\int \Big(
-f_Hh^{\mu\nu}\frac{\delta\bar{\bar{\Gamma}}}{\delta h^{\mu\nu}}
+f_HH^{\mu\nu}\frac{\delta\bar{\bar{\Gamma}}}{\delta H^{\mu\nu}}
      -f_Y\varphi\frac{\delta\bar{\bar{\Gamma}}}{\delta \varphi}
         +f_YY\frac{\delta\bar{\bar{\Gamma}}}{\delta Y}&&\\
+f_Lc\frac{\delta\bar{\bar{\Gamma}}}{\delta c}
-f_LL\frac{\delta\bar{\bar{\Gamma}}}{\delta L} \Big)
+\frac{\partial\bar{\bar{\Gamma}}}{\partial \alpha_0}&&=0.
\end{eqnarray}
(\ref{ntrm2}) corresponds to (\ref{fbrst}), hence we know that the general 
solution (of the linear case) is given by  
\begin{eqnarray}\label{frt}
\bar{\bar{\Gamma}}&=&
    \hat{c}_3\kappa^{-2}\int\sqrt{-g}R(z_1(\alpha_0)h)
   +\hat{c}_1\int\sqrt{-g}R^{\mu\nu}R_{\mu\nu}(z_1(\alpha_0)h)
   +\hat{c}_2\int\sqrt{-g}R^2(z_1(\alpha_0)h)\nonumber\\
&&+\hat{c}_{\rm R}\int(-g)^{1/4}\,\frac{1}{2}\varphi^2(z_\varphi(\alpha_0))R(z_1(\alpha_0)h)\\
	&& +\hat{c}_\varphi\int(-g)^{1/4}\frac{1}{2}g^{\mu\nu}
	   (\mathcal{D}_\mu\varphi\mathcal{D}_\nu\varphi)(z_\varphi(\alpha_0))
  +\hat{c}_\lambda\int(-\frac{\lambda}{4!}\varphi^4
						 (z_\varphi(\alpha_0)))\\
 &&+\hat{c}_H\int(\kappa H_{\mu\nu}(\frac{y(\alpha_0)}{z_1(\alpha_0)}
                  (-\partial^\mu c^\nu-\partial^\nu c^\mu)
                  -y(\alpha_0)(\partial_\lambda c^\mu h^{\lambda\nu}
                  -c^\lambda\partial_\lambda h^{\mu\nu}
                  +c^\lambda\partial_\lambda  h^{\mu\nu})) \\
   &&\qquad+t(\alpha_0)Y(c^\rho\partial_\rho\varphi
			 +\frac{1}{4}\partial_\rho c^\rho\varphi)
	      -\kappa y(\alpha_0) L_\rho\partial^\lambda c^\rho)     . \nonumber
\end{eqnarray}
(\ref{frt}) inserted into (\ref{ntrm3}) implies after some calculations that all $\hat{c}$ 
are independent of $\alpha_0$, whereas the functions $f_{H,L}$
satisfy the relations  
\be\label{rsfrt}
     \partial_{\alpha_{0}}z_1=f_H z_1 \qquad
     \partial_{\alpha_{0}}y=-f_L y  \qquad
     \partial_{\alpha_0}z_\varphi=f_\varphi z_\varphi
\ee
All parameters $\hat{c}$ can therefore be fixed by normalization conditions
independent of $\alpha_0$. Since we shall work in Landau gauge,
$\alpha_0=0$, the functions $f_H,f_L,f_Y$ will be independent of
$\alpha_0$, as well as $z_1,z_\varphi,y$, hence numbers. 

\subsection{Normalization conditions I}
In the tree approximation as studied in this section the free parameters of 
the model can be prescribed by the following conditions  
\begin{eqnarray}%\label{trnorm}
	\frac{\partial}{\partial p^2}\,\gamma^{(2)}_{\rm TT}|_{p^2=0}&
	=&c_3\kappa^{-2}\qquad ({\rm coupling\,\, constant})\label{trnorm3}\\
\frac{\partial}{\partial p^2}\frac{\partial}{\partial p^2}\,
\gamma^{(2)}_{\rm TT}&
	=&-2c_1\qquad({\rm coupling\,\, constant})\label{trnorm1}\\
\frac{\partial}{\partial p^2}\frac{\partial}{\partial p^2}\,\gamma^{(0)}_{\rm TT}
	&=&2(3c_2+c_1)\qquad (\rm{coupling\,\,constant})\label{trnorm2}\\
	\Gamma_{h^{\mu\nu}}&=&-\eta_{\mu\nu}c_0\doteq0\qquad(\rm{coupling\,\,constant})\label{trnorm0}\\
  \frac{\partial^2}{\partial p^\rho\partial p^\sigma}\Gamma_{\varphi\varphi h^{\mu\nu}}
	&=&\frac{c_{\rm R}}{2}(\delta^\mu_\rho\delta^\nu_\sigma+\delta^\mu_\sigma\delta^\nu_\rho)\qquad ({\rm coupling\,\, constant})\\
    \Gamma_{\varphi\varphi\varphi\varphi}&=&-\lambda \qquad ({\rm coupling\,\,constant})\\
 \frac{\partial}{\partial p_\sigma}\Gamma_{K^{\mu\nu}c_\rho}&=&
                                -i\kappa(\eta^{\mu\sigma}\delta^\nu_\rho
	                          +\eta^{\nu\sigma}\delta^\mu_\rho
	-\eta^{\mu\nu}\delta^\sigma_\rho)\qquad ({\rm amplitude\,\,of\,\,h\,\,and\,\, K})\label{trnorm4}\\
	\frac{\partial}{\partial p^\lambda}\Gamma_{L_\rho c^\sigma c^\tau}&=&
				-i\kappa(\delta^\rho_\sigma\eta_{\lambda\tau}
-\delta^\rho_\tau\eta_{\lambda\sigma})\qquad({\rm amplitude\,\,of\,\,c\,\,and\,\,L})\label{trnorm5}\\
\frac{\partial}{\partial p^\lambda}\Gamma_{Yc^\rho\varphi}&=&
  -i\delta_\lambda^\rho 
	      \qquad({\rm amplitude\,\, of}\,\,\varphi\,\,{\rm and\,\, Y})
\end{eqnarray}
Imposing the $b$-equation of motion (\ref{beq}) fixes $\alpha_0$ and the
$b$-amplitude.
It is worth mentioning that the $c_3$-contribution to $\gamma^{(0)}_{\rm TT}$ is an implication of the invariance under $\brsts_1 h$, hence must not be postulated via some normalization condition.\\

\section{Renormalization}
%file Renormalization.tex\\
At first we have to specify the perturbative expansion in which we would
like to treat the model. Due to the vanishing canonical dimension of the
field $h^{\mu\nu}$ we have to expand in the number of this field. Second
we expand as usual in the number of loops. Next we have to choose a
renormalization scheme in order to cope with the divergences of the loop
diagrams. We shall use the Bogoliubov-Parasiuk-Hepp-Zimmermann-Lowenstein
(BPHZL) scheme  \cite{LE}
which is based on momentum subtractions and an auxiliary
mass $M$ in order to avoid spurious infrared divergences which otherwise
would be introduced by the subtractions at vanishing momenta when dealing 
with massless propagators.\\
The key ingredients of this scheme are the subtraction operator
acting on one-particle-irreducible diagrams (1PI) and the forest formula 
which organizes the subtractions.  The subtraction operator reads
\be\label{sbtr}
(1-\tau_\gamma)=(1-t^{\rho(\gamma)-1}_{p^\gamma(s^\gamma-1)})
                (1-t^{\delta(\gamma)}_{p^\gamma s^\gamma}).
\ee
Here $t^d_{x_1...x_n}$ denotes the Taylor series about $x_i=0$ to order
$d$ if $d\ge 0$ or $0$ if $d<0$. $\gamma$ denotes a 1PI diagram, $p^\gamma$
refers to its external momenta, and $s^\gamma$ to an auxiliary subtraction
variable to be introduced. 
$\rho(\gamma)$ and  $\delta(\gamma)$ are the infrared and ultraviolet subtraction degrees of $\gamma$, respectively.
Those will be specified below. As far as the forest formula is concerned we 
refer to the literature (cf. \cite{LE}).
For later use we note that
\be\label{rmss}%removalssubtractions
(1-\tau_\gamma)=(1-t^{\delta(\gamma)}_{p^\gamma})
			\qquad {\rm for}\quad \rho(\gamma)=\delta(\gamma)+1 .
\ee

\subsection{Auxiliary mass}
In the BPHZ subtraction scheme one removes UV divergences by suitable subtractions at vanishing external momenta. In the massless case those would introduce artificial (off-shell) IR divergences. Hence in an extension, the BPHZL scheme, one introduces an auxiliary mass term of type $M^2(s-1)^2$ for every massless propagator. Subtractions with respect to $p,s$ performed at $p=0,s=0$ take care of the UV divergences. Subtractions with respect to $p,s-1$ thereafter establish correct normalizations for guaranteeing
 poles at $p=0$ and vanishing of three-point functions (of massless fields) at  $p=0$ .\\
For the massless scalar field this auxiliary mass term can simply be taken to read
\be\label{xmssf}
\frac{1}{2}\int M^2(s-1)^2\varphi^2.
\ee
When trying to introduce such an auxiliary mass term for the massless pole in the double pole propagators one encounters difficulties. Neither with a naive $hh$-term nor with a Fierz-Pauli type mass term can one invert $\Gamma_{hh}$ to propagators 
$G_{hh}$ such that the Lagrange multiplier field $b_\rho$ remains non-propagating. But its propagation would prevent its use in the quartet formalism of \cite{KuOj}. A glance at the propagators (\ref{bosprop}) and the coefficients 
$\gamma^{({\rm r})}_{\rm KL}$, (\ref{coffs}) suggests to replace
the overall factor $p^2$ in the $\gamma$’s by 
\be\label{axms}
p^2 - m^2 \equiv p^2-M^2(s-1)^2 .
\ee
Here $m^2$ denotes the auxiliary mass contribution.
This \emph{Push} in $p^2$ still maintains restricted invariance, i.e.\ under 
$\brsts_0 h$, (see Sect.\ 5.2), and is 
fairly easy to carry along as we shall see.\\
Accepting this change of vertices and propagators one has to analyze in some
detail what it implies. For the propagators it is clear that the pole at $p^2=0$
is shifted, as desired to a pole at $p^2=m^2$. It affects not only the invariant
parts, but also the gauge fixing dependent propagators $\langle bh \rangle$ and $\langle \bar{c}c \rangle$.
This can be seen when performing Push in $\Gamma$ and having a look at the 
inversion equations. The $\gamma$'s (\ref{coffs}) then read
\begin{eqnarray}\label{mcoffs}
	\gamma^{(2)}_{TT} &=&-(p^2-m^2)(c_1p^2-c_3\kappa^{-2})\label{gmps1}\\
\gamma^{(0)}_{TT} &=&(p^2-m^2)((3c_2+c_1)p^2-2c_3\kappa^{-2})\label{gmps2}\\ 
 \gamma^{(1)}_{SS}&=& \gamma^{(0)}_{WW}=\gamma^{(0)}_{TW}=\gamma^{(0)}_{WT}
	             =0 .
\end{eqnarray}
In the inversion equations one has products of $\gamma^{(r)}_{KL}$ with
its direct counterpart $\langle hh \rangle^{(r)}_{KL}$, such that this change is not a change there.\\
For gauge fixing terms we find the effect of Push as follows
\begin{eqnarray}\label{mfg}
\Gamma^{hb}_{\mu\nu\rho}G^{bh}&=&\frac{i}{2\kappa}(\eta_{\rho\mu}p_\nu+\eta_{\rho\nu} p_\mu)\frac{\kappa}{p^2}(p^\mu\theta^{\nu\rho}+p^\nu\theta^{\mu\rho}+p^\rho\omega^{\mu\nu})                              \qquad{\rm (local)}\\
&=&\frac{i}{2\kappa}(\eta_{\rho\mu}p_\nu+\eta_{\rho\nu} p_\mu)
       \frac{p^2}{p^2}\frac{\kappa}{p^2}(p^\mu\theta^{\nu\rho}+p^\nu
       \theta^{\mu\rho}+p^\rho\omega^{\mu\nu})
       \qquad{\rm (local)}                                          \\
		&\stackrel{\rm Push}{\rightarrow}&\frac{i}{2\kappa}(\eta_{\rho\mu}p_\nu
+\eta_{\rho\nu}p_\mu)\frac{p^2-m^2}{p^2}\frac{\kappa}{p^2-m^2}
(\theta^{\rho\mu}p^\nu+\theta^{\rho\nu} p^\mu
                                      +p^\rho\omega^{\mu\nu})\\
\Rightarrow\Gamma(m^2)^{hb}_{\mu\nu\rho}&=&\frac{-im^2}{2\kappa p^2}(\eta_{\rho\mu}p_\nu+\eta_{\rho\nu} p_\mu)   \qquad({\rm non-local}),\label{gmgf}\\
\Rightarrow G^{bh}_{\rho\mu\nu}&=&\frac{\kappa}{p^2-m^2}(p_\mu\theta_{\rho\nu}
+ p_\nu\theta_{\rho\mu}+p_\rho\omega_{\mu\nu})\qquad({\rm massive\,\, propagator})
\end{eqnarray}
i.e.\ there appears an additional term in $\Gamma^{hb}$ and the $\langle bh \rangle$-propagator
becomes massive (with the auxiliary mass). 
In $x$-space complete gauge fixing term reads
\begin{eqnarray}\label{cmgf}
	\Gamma_{\rm{gf}}&=&-\frac{1}{2\kappa}\int dxdy\, h^{\mu\nu}(x)
(\partial_\mu b_\nu+\partial_\nu b_\mu)(y)\Big\lbrace\delta(x-y)
+\frac{m^2}{(x-y)^2}\Big\rbrace -\frac{\alpha_0}{2}\int\eta^{\mu\nu} b_\mu b_\nu\nonumber\\
	            &=&-\frac{1}{2\kappa}\int dxdy\, h^{\mu\nu}(x)
	(\partial_\mu b_\nu+\partial_\nu b_\mu)(y)\Big\lbrace\big( \frac{\Box}{4\pi^2} + m^2 \big)
\frac{1}{(x-y)^2}\Big\rbrace -\frac{\alpha_0}{2}\int\eta^{\mu\nu} b_\mu b_\nu.
\end{eqnarray}
A suitable Faddeev-Popov (FP) term is then
\begin{eqnarray}\label{FaPo}
	\Gamma_{\phi\pi}&=&-\frac{1}{2}\int dxdy\, D^{\mu\nu}_\rho c^\rho(x)
(\partial_\mu \bar{c}_\nu+\partial_\nu \bar{c}_\mu)(y)\lbrace\delta(x-y)
	               +\frac{m^2}{(x-y)^2}\rbrace\nonumber\\
	&=&-\frac{1}{2}\int dxdy\, D^{\mu\nu}_\rho c^\rho(x)
(\partial_\mu \bar{c}_\nu
	+\partial_\nu \bar{c}_\mu)(y)\lbrace\big( \frac{\Box}{4\pi^2} + m^2 \big)\frac{1}{(x-y)^2}\rbrace ,
\end{eqnarray}
because it maintains the BRST-doublet structure within the gauge fixing procedure.\\
A comment to the ``non-local'' terms is in order. Our writing is
symbolic shorthand in order to have a simple handling of these terms. Using the 
explicit form of $\brsts_0 h$ and integration by parts one may observe that the actual non-local part is of projector type in terms of differential operators -- quite in line with its first appearance in $p$-space. 
There the projectors lead formally to direction dependent integrals. However Zimmermann's $\varepsilon$, introduced as
\begin{align}
	p^2 \to p^2+i\epsilon({\mathbf p}^2+M^2(s-1)^2) \, , \label{psl}
\end{align}
guarantees absolute convergence, hence no serious problem will arise once we
have reliable power counting and appropriate correct subtractions.
In the limit $\epsilon\to 0$ its contribution vanishes.\\ 
We therefore discuss in the next subsection power counting and convergence with positive
outcome, and return thereafter to a discussion of the $m^2$-dependent terms.
Before starting with the presentation of power counting we have to have a look at the basis of naively symmetric insertions
once we have introduced an auxiliary mass term. 
Obviously we can introduce the following \emph{Shift}
\be\label{ivce}
\int\sqrt{-g}c_3\kappa^{-2}R \to \int\sqrt{-g}(c_{30}\kappa^{-2}
+c_{31}\kappa^{-1}m+c_{32}\frac{1}{2}m^2)R.
\ee
In the tree approximation these terms are invariant (and for 
$s=1$ reduce to the original term), but in higher orders they represent new and independent elements in the basis of symmetric normal products
with $\delta=\rho=4$. 
So, we have to carry them along as vertices when studying power counting.

\subsection{Power counting and convergence}
In the Landau gauge, $\alpha_0=0$, the only non-vanishing propagators are
the following one's:  
\begin{eqnarray}\label{sprops}
	\langle hh \rangle^{(2)}_{TT}&=&\frac{i}{(p^2-m^2)c_1( p^2-\frac{c_3\kappa^{-2}}
	{c_1})}\\
	\langle hh \rangle^{(0)}_{TT}&=&\frac{i}{(p^2-m^2)(3c_2+c_1)(p^2-\frac{2c_3\kappa^{-2}}{(3c_2+c_1)})}\\
	\langle b_\rho h_{\mu\nu} \rangle&=&\frac{1}{p^2-m^2}(p_\mu\theta_{\nu\rho}
	+p_\nu\theta_{\mu\rho}+p_\rho\omega_{\mu\nu})\\
	\langle \bar{c}_\rho c_\sigma  \rangle&=&
-i\big(\theta_{\rho\sigma}+\frac{1}{2}\omega_{\rho\sigma} \big)\frac{1}{p^2-m^2}\\
	\langle \varphi\varphi \rangle&=&\frac{i}{p^2-m^2}
\end{eqnarray}
In addition to $m=M(s-1)$ one needs also Zimmermann's $\varepsilon$-prescription (\ref{psl}). This will guarantee absolute convergence of diagrams, once power counting
is established and subtractions are correctly performed.\\
Important note: in all formulas to follow in this section
the replacement of $c_3$ by the sum given in (\ref{ivce}) is to be understood. Relevant for power counting arguments is never
a coefficient in front of a vertex, but the number of lines and derivatives at the vertex and its associated subtraction degree.
The $\langle bh \rangle$ propagator will be of no relevance for reasons spelled out after (\ref{znt2}). 

Power counting is based on ultraviolet (UV) and infrared (IR) degrees
of propagators and vertices. The upper degree $\overline{\rm deg}_{p,s}$ gives the asymptotic power for $p$ and $s$ tending to infinity; the lower degree 
$\underline{\rm deg}_{p,(s-1)}$ gives the asymptotic power for $p$ and $s-1$ tending to zero.
For propagators they read 
\begin{eqnarray}\label{dgpr}
	{\overline{\rm deg}}_{p,s}(\langle hh \rangle^{(2)}_{TT})&=&-4 \qquad{}
	{\underline{\rm deg}}_{p,s-1}(\langle hh \rangle^{(2)}_{TT})=-2 \\
	{\overline{\rm deg}}_{p,s}(\langle hh \rangle^{(0)}_{TT})&=&-4 \qquad
	{\underline{\rm deg}}_{p,s-1}(\langle hh \rangle^{(0)}_{TT})=-2 \\
        {\overline{\rm deg}}_{p,s}(\langle \bar{c}c \rangle)&=&
	 {\underline{\rm deg}}_{p,s-1}(\langle \bar{c}c \rangle)=-2 \\
        {\overline{\rm deg}}_{p,s}(\langle \varphi\varphi \rangle)&=&
	           {\underline{\rm deg}}_{p,s-1}(\langle \varphi\varphi \rangle)=-2 .
\end{eqnarray}
As shorthand we write also 
$\overline{\rm deg}\equiv \overline{D}_L$ and $\underline{\rm deg}\equiv \underline{D}_L$.
The degrees of the vertices thus have the values
\begin{eqnarray}\label{dgve}
\overline{D}_{V^{(c_1)}}&=&\overline{D}_{V^{(c_2)}}=4, 
	 \quad \overline{D}_{V^{(c_3)}}=
	 \overline{D}_{V^{(\phi\pi)}}= 
	 \overline{D}_{V^{(kin)}}=
	 \overline{D}_{V^{(c_{\rm R})}}=2,\quad
	 \overline{D}_{V^{\varphi}}=0\\
\underline{D}_{V^{(c_1)}}&=&\underline{D}_{V^{(c_2)}}=4,\quad 
	 \underline{D}_{V^{(c_3)}}=
	 \underline{D}_{V^{(\phi\pi)}}=
	 \underline{D}_{V^{(kin)}}=
	 \underline{D}_{V^{(c_{\rm R})}}=2, \quad
	 \underline{D}_{V^{\varphi}}=0
\end{eqnarray}
In addition to the vertices of the EH theory we have to take into account
\be\label{IIvrtcs}
V^{(c_R)}=\frac{c_R}{2}\int(-g)^{1/4}\varphi^2R \quad
V^{(kin)}=\frac{1}{2}\int(-g)^{1/4}g^{\mu\nu}\mathcal{D}_\mu\varphi
                 \mathcal{D}_\nu\varphi \quad
V^{(\varphi)}=-\frac{\lambda}{4!}\int\varphi^4
\ee
Let us now consider a one-particle-irreducible (1PI) diagram $\gamma$ with $m$ 
loops, $I_{ab}$ internal lines, $a,b = h,c,\bar{c}$, and $V$
vertices of type $V\in\{c_1,c_2,c_3,\phi\pi,c_{\rm R},kin,\varphi\}$ or insertions $Q_i$ as well as $N$ amputated 
external lines. In the subsequent considerations a more detailed notation is useful: $N_a$ are of type $\Phi_a$,
$n_{ai}$ are of type $a$ and are attached to the $i^{\rm th}$ vertex. Then 
with $Q_i$
\be\label{ins} 
Q_i(x)=(\frac{\partial}{\partial x})^{|\mu_i|}\prod_a(\Phi_a^{c_{ai}}(x)) ,
\ee
we first find for the UV- and IR-degrees of $\gamma$ 
\begin{align}\label{gmmadeg}
d(\gamma)&=4m(\gamma)+\sum_{V\in\gamma}\overline{D}_V+\sum_{L\in\gamma}\overline{D}_L\\
      &=4m(\gamma) +4V^{(c_1,c_2)}+2(V^{(c_3)}+V^{(\phi\pi)}+V^{(c_R)}+V^{(kin)})
	               -4I_{hh}-2(I_{c\bar{c}}+I_{\varphi\varphi}),\\
r(\gamma)&=4m(\gamma)+\sum_{V\in\gamma}\underline{D}_V
	                             +\sum_{L\in\gamma}\underline{D}_L\\
   &=4m(\gamma)+4V^{(c_1,c_2)}+2(V^{(c_3)}+V^{(\phi\pi)}+V^{(c_R)}+V^{(kin)})
	 -2(I_{hh}+I_{\bar{c}c}+I_{\varphi\varphi}) .
\end{align}
The topological relations
\begin{eqnarray}\label{topform}
	m&=&I-V+1\\
	N_a&=&\sum_i n_{ai} 
	\qquad 2I_{aa}=\sum_i(c_{a i}-n_{ai})=\sum_ic_{ai}-N_a
\end{eqnarray}
permit to rewrite these degrees as
\begin{eqnarray}\label{uvirdeg}
	d(\gamma)&=&4+\sum_{V\in\gamma}(\overline{D}_V-4)
	                      +\sum_{L\in\gamma}(\overline{D}_L+4)\\
	d(\gamma)&=&4-(N_c+N_{\bar{c}})-2V^{(c_3)}-N_{\varphi}\\ 
	r(\gamma)&=&4+\sum_{V\in\gamma}(\underline{D}_V-4)
			     +\sum_{L\in\gamma}(\underline{D}_L+4)\\
	r(\gamma)&=&4-2V^{(c_3)}+2I_{hh}-(N_c+N_{\bar{c}}+N_\varphi).
\end{eqnarray}
The aim is now to associate subtraction degrees to them which are independent
of the detailed structure of the respective diagrams. We chose
\be\label{subtrdeg}
\delta(\gamma) = 4-N_\varphi \qquad \rho(\gamma) = 4-N_\varphi,
\ee
\noindent
i.e.\ make explicit the dependence on the number of external legs $N_\varphi$,
because these are the standard degrees for a massless scalar field in the case
where no $h$-fields contribute.\\ 
(A side remark: In a model with additional massless vector and massless spinor 
fields and no
internal symmetry breaking one would add $-N_V-3/2 N_\psi$ in the rhs of 
(\ref{subtrdeg}). Hence the standard model before symmetry breaking would
be covered. Symmetry breaking would however require model dependent, 
``dedicated'' degree prescriptions \cite{EK98}.)\\
We now have to check, that Lowenstein's conditions \cite{Lo} 
are still satisfied. The first one reads
\be\label{c1} 
	\delta(\gamma)= d(\gamma)+b(\gamma) 
			\quad\mbox{and}\quad \rho(\gamma)=  r(\gamma)-c(\gamma) \tag{C1}
\ee
with $b(\gamma)$ and $c(\gamma)$ being non-negative integers. 
$b(\gamma)\ge 0$ is obviously satisfied, but for 
\be\label{irc}
c(\gamma)=r(\gamma)-\rho(\gamma)=-2V^{c_3}+I_{hh}-N_c
\ee 
we have to convince ourselves that it is greater or equal to zero. Hence we need the more detailed information given by the line balances
\begin{align}\label{linetop}
	2I_{hh}&= \sum_{i\in \gamma}(c_{h,i}-n_{h,i}) 
	      =  \sum_{i\in \gamma}(c_{h,i})-N_h 
		  \quad i\in\{V^{(c_1)},V^{(c_2}), V^{(c_3)},V^{(\phi\pi)},
		  V^{(c_R)},V^{(kin)},V^{(\varphi)}\}\\
2I_{c\bar{c}}&=\sum_{i\in\phi\pi}(c_{c,i}-n_{c,i})
              = \sum_{i\in\phi\pi}c_{c,i}-N_{c} \\
2I_{\varphi\varphi}&=\sum_{i=c_R,k,\varphi}(c_{\varphi,i}-n_{\varphi,i})=
                     \sum_{i=c_R,k,\varphi}(c_{\varphi,i}-N_\varphi).              
\end{align}
We find
\begin{align}\label{cg}
c(\gamma)&=\sum_{i\in c_1,c_2}(c_{h,i}-n_{h,i})
           +\sum_{i\in c_3}(c_{h,i}-n_{h,i}-2)
           +\sum_{i\in \phi\pi}(c_{\tilde{c},i}-n_{\tilde{c},i}-2)
           +\sum_{i\in\phi\pi}(1-n_{h,\phi\pi})\\
         &+\sum_{i\in V^{(c_R)}}(c_{h,V^{(c_R)}}-n_{h,V^{(c_R)}})
           +\sum_{i\in V^{(kin)}}(c_{h,V^{(kin)}}-n_{h,V^{(kin)}})
         +\sum_{i\in V^{(\varphi)}}(c_{h,V^{(\varphi)}}-n_{h,V^{(\varphi)}})
\end{align}
If the vertex $i$ in question is not present in $\gamma$, the respective brackets
just vanish. If this vertex is present in $\gamma$, then (first line) 
$(c_{h,i}-n_{h,i})\ge 2$ and $(c_{h,i}-n_{h,i}-2)\ge 0$ -- both for 1PI $\gamma$.
Since $c_{\tilde{c},\phi\pi}=2$  the third bracket combines with the
fourth such that their sum is $\ge 0$ -- again for 1PI $\gamma$ -- we find two cases:
either $n_{h,i_0}=1$ at vertex $i_0$ s.t. $n_{\tilde{c},i_0}=0$ (otherwise 
$\gamma$ is not 1PI)
or $n_{h,i_0}=0$ at vertex $i_0$ s.t. $+1$ from here and from 
$n_{\tilde{c},i_0}$ at most 1, i.e. $-1$ in the sum (otherwise $\gamma$ is not 1PI), which together is $0$, i.e. non-negative. This refers to the old result 
in I. In the second line every bracket is grater or equal to zero if the respective vertex is in the diagram. Hence equations (\ref{c1}) are valid.\\

The next requirements refer to reduced diagrams 
$\bar{\Lambda}=\Lambda/\lambda_1,...\lambda_n$, which are obtained from
$\Lambda$
by contracting mutually disjoint, non-trivial 1PI subdiagrams $\lambda_i$
to points (reduced vertices) $V(\lambda_i)$ assigning (for the sake of power 
counting) the unit polynomial of momenta to each $V(\lambda_i)$. For 1PI 
$\gamma$ one has the relations
\begin{eqnarray}\label{subdeg}
	d(\gamma)&=&d(\gamma/\lambda_1...\lambda_n)+\sum_{i=1}^{n}d(\lambda_i)\\
	r(\gamma)&=&r(\gamma/\lambda_1...\lambda_n)+\sum_{i=1}^{n}r(\lambda_i).
\end{eqnarray}
Their analoga are also valid for connected diagams.
Now one can formulate further conditions for convergence, i.e.\ 
\begin{align}\label{c2}
	\delta(\gamma)&\ge d(\gamma/\lambda_1...\lambda_n)
	          +\sum_{i=1}^{n}\delta(\lambda_i) \tag{C2}\\
	\rho(\gamma)&\le r(\gamma/\lambda_1...\lambda_n)
		  +\sum_{i=1}^{n}\rho(\lambda_i) \tag{C3} \label{c3}\\
	\rho(\gamma)&\le \delta(\gamma)+1 \tag{C4} \label{c4}
\end{align}
for arbitrary reduced 1PI subdiagrams $\gamma/\{\lambda_i\}$ of $\Gamma$.
In order to verify \eqref{c2} one just inserts the values for the respective 
degrees.
\begin{eqnarray}\label{cC2}
   \delta(\gamma)&=&4-N_\varphi(\gamma) \\
 \delta(\gamma_i)&=&4-N_\varphi(\gamma_i)\\
d(\gamma)&=&4-2V^{(c_3)}(\gamma)-(N_{\bar{c}}+N_c)(\gamma)-N_\varphi(\gamma)\\
d(\gamma_i)&=&4-2V^{(c_3)}(\gamma_i)-(N_{\bar{c}}+N_c)(\gamma_i)
	                                 -N_\varphi(\gamma_i)\\
d(\bar{\gamma})&=&4-2V^{(c_3)}(\bar{\gamma})-(N_{\bar{c}}+N_c)(\bar{\gamma})
	                  -N_\varphi(\bar{\gamma})-4n\\
d(\bar{\gamma})+\sum_i\delta(\gamma_i)&=&4-2V^{(c_3)}(\bar{\gamma})
	-(N_{\bar{c}}+N_c)(\bar{\gamma})-N_\varphi(\bar{\gamma})\\
\delta(\gamma)=4-N_\varphi(\gamma)&\ge& 4-2V^{(c_3)}(\bar{\gamma})
         -(N_{\bar{c}}+N_c)(\bar{\gamma})-N_\varphi(\bar{\gamma}) 
\end{eqnarray}
The last inequality was to be proved.

For the proof of (C3) one can use literally the proof of (C2)
in the opposite direction since $N_\varphi(\gamma)=N_\varphi(\bar{\gamma})$.\\
\eqref{c4} is satisfied by definition of $\rho,\delta$.\\

We can now refer to \cite{Lo}, (theorem 4) in which 
it is shown that these conditions being satisfied, Green's functions exist as
tempered distributions, whereas for non-exceptional momenta (Euclidean sense)
vertex functions exist as functions. Due to a theorem of Lowenstein and Speer
\cite{LS}
in the limit $\varepsilon \rightarrow 0$ Lorentz covariance is also satisfied.
An important improvement concerning Lorentz covariance has been provided by 
\cite{CL}. If one introduces Zimmermann's $\varepsilon$
via a change of metric 
$\eta_{\mu\nu} \to {\rm diag}(1,-(1-i\varepsilon),-(1-i\varepsilon), -(1-i\varepsilon) )$ in additon to multiplying each mass-square by 
$(1-i\varepsilon)$ then Lorentz covariance already holds for the rhs of ZI's before establishing the $\varepsilon\to 0$ limit. This is quite helpful for actual work with
ZI's.

The above proof of convergence refers to diagrams constructed out of vertices
with vanishing Faddeev-Popov (FP) charge. For installing the ST-identity in higher
orders one needs however diagrams which once contain the vertex $V^{(-)}$ of types
\begin{eqnarray}\label{fpm1}
\overline{D}(V^{(-)})=\left\{
	\begin{array}{ll}
3& {\rm for}\quad V^{(-)}\simeq \int c\,\partial\partial\partial\, h\cdots h\\  
5& {\rm for}\quad V^{(-)}\simeq \int c\,\partial\partial\partial\partial\partial\, 
	                                                   h\cdots h 
	\end{array}	\right. &
	\underline{D}(V^{(-)})=\overline{D}(V^{(-)}),
\end{eqnarray}
i.e.\ of FP-charge $-1$. Matter field contributions to $V^{(-)}$ have the form
\begin{eqnarray}\label{mmo}
\overline{D}(V^{(-)})=\left\{
	\begin{array}{ll}
5& {\rm for}\quad V^{(-)}\simeq \int c\,\partial\partial\partial\,\phi\phi\, h\cdots h\\  
5& {\rm for}\quad V^{(-)}\simeq \int c\,\partial\,\phi\phi\phi\phi \,
	                                                   h\cdots h 
	\end{array}	\right. &
	\underline{D}(V^{(-)})=\overline{D}(V^{(-)}),
\end{eqnarray}

The UV- and IR-degrees become resp.\
\begin{eqnarray}\label{gmmadeg2}
	d(\gamma)&=&4m(\gamma)+\sum_{V\in\gamma}\overline{D}_V
		       +\sum_{L\in\gamma}\overline{D}_L+\overline{D}_{V^{(-)}}  \\
r(\gamma)&=&4m(\gamma)+\sum_{V\in\gamma}\underline{D}_V
		       +\sum_{L\in\gamma}\underline{D}_L+\underline{D}_{V^{(-)}}  .
\end{eqnarray}
With (\ref{topform}) this results into $(V^{(-)}\in \gamma)$
\begin{eqnarray}\label{vld}
	d(\gamma)&=&4+\sum_{V\in \gamma}(\overline{D}_V-4)
		   +\sum_{L\in\gamma}(\overline{D}_L+4)\\
	 &=&4-(N_{\bar{c}}+N_c)-N_\varphi-2V^{(c_3)} +(\overline{D}_{V^{(-)}}-4)\\ 
	r(\gamma)&=&4+\sum_{V\in \gamma}(\underline{D}_V-4)
		   +\sum_{L\in\gamma}(\underline{D}_L+4)\\
 &=&4-N_\varphi-2V^{(c_3)}-2V^{(\phi\pi)}+(\underline{D}_{V^{(-)}}-4)+2I_{hh}
                                                             +2I_{c\bar{c}}
\end{eqnarray}
As subtractions degrees we define
\begin{eqnarray}\label{sdr2}
	\delta(\gamma)&=d(\gamma)-N_\varphi
	+\left\{\begin{array}{ll}    
		0& {\rm if}\quad V^{(-)}\notin \gamma\\  
		1& {\rm if}\quad V^{(-)}\in \gamma
		\end{array}	\right. \\
	\rho(\gamma)&=r(\gamma)-N_\varphi
	+\left\{\begin{array}{ll}    
		0& {\rm if}\quad V^{(-)}\notin \gamma\\  
		1& {\rm if}\quad V^{(-)}\in \gamma .
		\end{array}	\right.
\end{eqnarray}
\be
{\rm i.e.}\qquad		\rho(\gamma)=\delta(\gamma) 
\ee
	The line balances read now
\begin{eqnarray}\label{lbcst}%linebalanceslavnovtaylor
	2I_{hh}&=&\sum_{i\in \gamma}(c_{h,i}-n_{h,i})\\ 
	      &=&\sum_{i\in \gamma}c_{h,i}-N_h \qquad 
	   i\in\{(c_1),(c_2),(c_3),(\phi\pi),(c_R),(kin),(\varphi),(-)\}\\
2I_{c\bar{c}}&=&\sum_{i\in\gamma}(c_{c,i}-n_{c,i})
               = \sum_{i\in\gamma}c_{c,i}-N_{c}
	  \qquad i\in\{(\phi\pi),(-)\} \\
2I_{\varphi\varphi}&=&\sum_{i\in\gamma}(c_{\varphi,c_i}-n_{\varphi,i})=
                                  \sum_{i\in\gamma}c_{\varphi,i}-N_\varphi
		       \qquad  i\in\{(c_{\rm R}),(kin),(\varphi),(-)\}	  
\end{eqnarray}
We just added the vertex $V^{(-)}$ to the original line balances (\ref{linetop}).\\
In order to verify \eqref{c1} we have to show that 
$b(\gamma)=\delta(\gamma)-d(\gamma)\ge 0$.
\begin{eqnarray}\label{c1dst}%C1dslavnovtaylor
	b(\gamma)&=&5-N_\varphi-d(\gamma)\\
     &=&5-4+2(V^{(c_3)}+V^{({\phi\pi})})
           -(\overline{D}^{V^{(-)}}-4)-2I_{c\bar{c}}\\
     &=&1+2V^{(c_3)}-1+\sum_{i\in\phi\pi}n_{\tilde{c},\phi\pi}-(1-n_{c,V^{(-)}})\\
     &=&2V^{(c_3)}+\sum_{i\in\phi\pi}n_{\tilde{c},\phi\pi}-(1-n_{c,V^{(-)}}) .
\end{eqnarray}
In the transition from first to second line the $\varphi$-line contributions cancel 
and we have used the line balance for $I_{c\bar{c}}$ (\ref{linetop})and chosen 
the more dangerous case $\overline{D}_{V^{(-)}}=5$.
If $n_{c,V^{(-)}}=0$, there must a $+1$ coming from the $\phi\pi$-sum, because
the FP-charge is conserved. This is true for the $V^{(-)}$-vertices depending on 
field $\varphi$ as well. Hence the inequality holds.\\

The control of 
\begin{align}\label{2c1r}
c(\gamma)=&r(\gamma)-\rho(\gamma)\\
	 =&4-2(V^{(c_3)}+V^{(\phi\pi)}+V^{(c_R)}+V^{(kin)})-4V^{(\varphi)}
	+2(I_{hh}+I_{c\bar{c}}+I_{\varphi\varphi})
               +(\underline{D}(V^{(-)})-4)-5\\
	=&-2V^{(c_3)} -2V^{(\phi\pi)}+2I_{hh}+2I_{c\bar{c}}
               +(\underline{D}(V^{(-)})-4)-1\\
	=&-2V^{(c_3)})-2V^{(\phi\pi)})+2I_{hh}+2I_{c\bar{c}}+ 
                \left\{\begin{array}{l} 
			-1 \,\,{\rm for}\,\,\underline{D}_{V^{(-)}}=3\\
			+1 \,\,{\rm for}\,\,\underline{D}_{V^{(-)}}=5 
	              \end{array} \right.\ge 0 .
\end{align}
is similar. In the difference $r(\gamma)-\rho(\gamma)$ the contribution from 
$\varphi$-lines drops out. The effect of the $\varphi$-contributions to 
$V^{(-)}$-vertices is taken along by their $\underline{D}$-value. The continuity of $c\bar{c}$-lines is maintained. Hence one
falls entirely back to the analysis of the pure EH-model, with the same result, that indeed $c(\gamma)$ is non-negative.\\

When checking (C2) and (C3) we encounter the same situation: the 
$\varphi$-lines
and their contributions just go through the estimates since 
$N_\varphi(\gamma)=N_\varphi(\bar{\gamma})$ and the decisive steps are those of the underlying pure EH-model: it is the conservation of the 
Faddeev-Popov-charge, i.e.\ the contingency of the $c\bar{c}$-lines which is 
relevant and not the specific form of the vertices $V^{(-)}$ depending on 
$\varphi$.\\
From here on the discussion of [PS21] can be literally taken over with the
result that one has convergence for all insertions needed in the sequel.
Again, the limiting case $\rho(\gamma)=\delta(\gamma)+1$ can not
be used: The subtractions must be performed with the above indicated
degreees $\delta(\gamma),\rho(\gamma)$.\\

\subsection{Slavnov-Taylor identity}
The ST identity which we have to establish to higher orders takes the same form 
as in tree approximation, (\ref{fbrst}), supplemented however
by the $m^2$-dependent gauge fixing, (\ref{cmgf}), and Faddeev-Popov-terms, 
(\ref{FaPo}), i.e.
\be\label{2fbrst}
\mathcal{S}(\Gamma)\equiv
\int\Big(\frac{\delta\Gamma}{\delta{K}}\frac{\delta\Gamma}{\delta h}
+\frac{\delta\Gamma}{\delta L}\frac{\delta\Gamma}{\delta c}
+\frac{\delta\Gamma}{\delta Y}\frac{\delta\Gamma}{\delta c}
+b\frac{\delta\Gamma}{\delta\bar{c} }\Big)=0
\ee
\begin{eqnarray}
\label{2cmgf}
	\Gamma_{\rm gf}&=&-\frac{1}{2\kappa}\int dxdy\, h^{\mu\nu}(x)
	(\partial_\mu b_\nu+\partial_\nu b_\mu)(y)\Big\lbrace\big( \frac{\Box}{4\pi^2} + m^2 \big)
\frac{1}{(x-y)^2}\Big\rbrace \\
 &&-\int\frac{\alpha_0}{2}\eta^{\mu\nu} b_\mu b_\nu\\
\label{2FaPo}
	\Gamma_{\phi\pi}&=&-\frac{1}{2}\int dxdy\, \brsts\,h^{\mu\nu}(x)
(\partial_\mu \bar{c}_\nu
	+\partial_\nu \bar{c}_\mu)(y)\Big\lbrace\big( \frac{\Box}{4\pi^2} + m^2 \big)\frac{1}{(x-y)^2}\Big\rbrace .
\end{eqnarray}
The $b,\bar{c}$-field equations of motion take now the form 
\begin{eqnarray}
\label{2beq}
\frac{\delta \Gamma}{\delta b^\rho}&=&\kappa^{-1}\int 
                              dy\,\partial^\mu h_{\mu\rho}(y)
             \Big\lbrace\big( \frac{\Box}{4\pi^2} + m^2 \big)\frac{1}{(x-y)^2}\Big\rbrace-\alpha_0 b_\rho\\
\label{2ghe}
	\frac{\delta\Gamma}{\delta \bar{c}_\rho(x)}&=&
	-\int dy\, \kappa^{-1} \partial_\lambda\frac{\delta\Gamma}{\delta K_{\lambda\rho}(y)} 
             \Big\lbrace\big( \frac{\Box}{4\pi^2} + m^2 \big)\frac{1}{(x-y)^2}\Big\rbrace .
\end{eqnarray}
Again the $b$-field equation can be integrated trivially back to (\ref{2cmgf}) and 
therefor the functional $\bar{\Gamma}$ be introduced as in the tree approximation
\be\label{2Gmmbr}
\Gamma = \Gamma_{\rm gf}+\bar{\Gamma} .
\ee
(\ref{rstc}) is changed into
\be\label{2rstc}
\kappa^{-1}\int dy\,
           \partial_\lambda\frac{\delta\bar{\Gamma}}{\delta K_{\mu\lambda}(y)}
			      \Big\lbrace\big( \frac{\Box}{4\pi^2} + m^2 \big)\frac{1}{(x-y)^2}\Big\rbrace
+\frac{\delta\bar{\Gamma}}{\delta\bar{c}_\mu} =0,
\ee
whereas (\ref{sceH}) becomes
\be\label{2sceH}
H_{\mu\nu}(x)=K_{\mu\nu}(x)
+\frac{1}{2}\int dy\,(\partial_\mu\bar{c}_\nu+\partial_\nu\bar{c}_\mu)(y)
   \Big\lbrace\big( \frac{\Box}{4\pi^2} + m^2 \big)\frac{1}{(x-y)^2}\Big\rbrace .
\ee
The relations (\ref{brGm}) are unchanged:
\begin{eqnarray}\label{2brGm}
\mathcal{S}(\Gamma)&=&\frac{1}{2}\mathcal{B}_{\bar{\Gamma}}\bar{\Gamma}=0\\
	\mathcal{B}_{\bar{\Gamma}}&\equiv& 
	\int\Big(
  \frac{\delta\bar{\Gamma}}{\delta H}\frac{\delta}{\delta h}
+ \frac{\delta\bar{\Gamma}}{\delta h}\frac{\delta}{\delta H}
+ \frac{\delta\bar{\Gamma}}{\delta Y}\frac{\delta}{\delta\varphi}
+ \frac{\delta\bar{\Gamma}}{\delta\varphi}\frac{\delta}{\delta Y} 
+ \frac{\delta\bar{\Gamma}}{\delta L}\frac{\delta}{\delta c}
+ \frac{\delta\bar{\Gamma}}{\delta c}\frac{\delta}{\delta L}\Big) .
\end{eqnarray}

In the BPHZL renormalization scheme the starting point for establishing equations
like the above one's to all orders is a $\Gamma_{\rm eff}$ with which one calculates accordingly subtracted Feynman diagrams. 
Here we choose
\be\label{Gmmff}
\Gamma_{\rm eff}=\Gamma^{\rm class}_{\rm inv} +\Gamma_{\rm gf}+\Gamma_{\phi\pi}
                                 +\Gamma_{\rm e.f.}+\Gamma_{\rm ct} .
\ee
In addition to (\ref{ivc}),(\ref{clssct}),(\ref{2cmgf}), and (\ref{2FaPo})
one has to take into account the changes caused by the auxiliary mass term
(\ref{axms}) in (\ref{gmps1}) and (\ref{gmps2}).
$\Gamma_{\rm ct}$ will collect counterterms as needed. All these expressions are 
to be understood as normal products, i.e.\ insertions into Green
functions with power counting degrees $\delta=\rho=4$.

Starting from $Z$, the generating functional for general Green functions, 
and from the definition of $\mathcal{S}$ in (\ref{sma}) we
postulate 
\be\label{Zbrst}
\mathcal{S}Z=0.
\ee
Then the action principle yields
\be\label{acZbrst}
\mathcal{S}Z=\Delta_Z\cdot Z= \Delta_Z +O(\hbar \Delta_Z),
\ee
where $\Delta_Z\equiv[\Delta_Z]^5_5$ is an integrated insertion with 
$Q_{\phi\pi}(\Delta_Z)=+1$.
Again, by invoking the action principle one can realize the $b$-field
equation of motion (\ref{2beq}), with (\ref{2rstc}), now on the renormalized
level, as a consequence of (\ref{2fbrst}). This admits (\ref{2brGm}) as a 
postulate and results into
\begin{eqnarray}\label{rgheq}
   \mathcal{S}(\Gamma)&=&\Delta\cdot\Gamma\\
\frac{1}{2}\mathcal{B}_{\bar{\Gamma}}\bar{\Gamma}&=&\Delta+O(\hbar\Delta) .
\end{eqnarray}
Here $\Delta\equiv [\Delta]_5^5$ with $Q_{\phi\pi}(\Delta)=+1$
does not dependent on $b$ and $\bar{c}$. These relations admit a cohomological 
treatment, since 
\be\label{cstc}
\mathcal{B}_{\bar{\Gamma}}\mathcal{B}_{\bar{\Gamma}}\bar{\Gamma} =0, \qquad
\mathcal{B}_{\bar{\Gamma}}\mathcal{B}_{\bar{\Gamma}}=0,
\ee
the latter being true as a necessary condition, if (\ref{2brGm}) is to be 
satisfied.
Since in the tree approximation (\ref{2brGm}) holds one has
\be\label{2cstc}
\brstb\Delta=0 \quad {\rm for} \quad \brstb\equiv \mathcal{B}_{\bar{\Gamma}_{\rm class}}
\qquad {\rm with} \quad \brstb^2=0
\ee
as the final consistency condition to be solved. 
The standard way to solve this cohomology problem is to list contributions to $\Delta$ by starting with terms depending on external fields and then those consisting of elementary fields only, i.e.
\be\label{chmlg}
\Delta= \int(K_{\mu\nu}\Delta^{\mu\nu}(h,c)+L_\rho\Delta^\rho(h,c)
             +Y\Delta^\varphi(\varphi,h,c)) +\Lambda(h,\varphi,c) .
\ee
All terms are insertions compatible with $[...]^5_5$
and $Q^{\phi\pi}=+1$. (Recall that 
$Q^{\phi\pi}(K)=Q^{\phi\pi}(Y)=-1,Q^{\phi\pi}(L)=-2$.)
In \cite{BBHa,BBHb} it is shown, that all these contributions eventually are $\brstb$-variations. 
This is true even for the $\Lambda$-term. 
This means that also in the present case no anomalies arise, the solution 
reads:
\be\label{fchgrv}
\Delta=\brstb \hat{\Delta}
\ee
with a $\hat{\Delta}$ which can be absorbed into $\Gamma_{\rm eff}$. 
In the quoted references the algebra leading to this result has been performed by using cohomological methods.
Without power counting and convergence and not within a concrete renormalization scheme, this represents a classical consideration.
In the present context we have, however, supplied it with ``analytic''
information, i.e.\ assured the existence of the relevant quantities as
insertions into existing Green functions. 
The result is thus that we have indeed a ST-identity which holds as inserted 
into general Green's functions of elementary fields, at non-exceptional momenta 
and $s=1$. \\

In principle one would now like to prove unitarity of the $S$-matrix along the 
lines given 
in the tree approximation. This is however not directly possible. 
If $c_1$ and $c_2$ are different from zero, the <hh>-propagator has two poles which
have to be disentangled and associated with respective fields. Then one has to
study carefully the norm properties of their particle states and give a 
prescription for handling the ``dangerous'' one's. This will be done in 
section 9 below.\\

\subsection{Normalization conditions II}
The normalization conditions (\ref{trnorm1})-(\ref{trnorm5}) have to be 
modified such 
that they are compatible with higher orders of perturbation theory: they
have to be taken at values in momentum space which are consistent with
the subtraction procedure.
They read  
\begin{eqnarray}\label{highnorm}
\frac{\partial}{\partial p^2}\,\gamma^{(2)}_{\rm TT\,|{\substack{p=0 \\ s=1} }}&=
				   &c_3\kappa^{-2}\\
\frac{\partial}{\partial p^2}\frac{\partial}{\partial p^2}\,
	\gamma^{(2)}_{\rm TT\,|{\substack{p^2=-\mu^2\\ s=1} }}&=&-2c_1\\
\frac{\partial}{\partial p^2}\frac{\partial}{\partial p^2}\,
	\gamma^{(0)}_{\rm TT\,|{\substack{p^2=-\mu^2 \\ s=1} }}
                                        	&=&2(3c_2+c_1)\\
	\Gamma_{h^{\mu\nu}} &=&-\eta_{\mu\nu}c_0=0\\
\frac{\partial^2}{\partial p^\rho\partial p^\sigma}
	\Gamma_{\varphi\varphi h^{\mu\nu}|{\substack{p=p_{sym}(-\mu^2) \\ s=1}}}
     &=&\frac{c}{2}(\delta^\mu_\rho\delta^\nu_\sigma+\delta^\mu_\sigma\delta^\nu_\rho)	\\		
	\Gamma_{\varphi\varphi\varphi\varphi|{\substack{p=p_{sym}(-\mu^2) \\ s=1}}}&=&-\lambda\\
\frac{\partial}{\partial p_\sigma}
	\Gamma_{K^{\mu\nu}c_\rho|{\substack{p^2=-\mu^2 \\ s=1} }}&=&
                                -i\kappa(\eta^{\mu\sigma}\delta^\nu_\rho
	                          +\eta^{\nu\sigma}\delta^\mu_\rho
		  -\eta^{\mu\nu}   \delta^\sigma_\rho)\label{highnorm1} \\
\frac{\partial}{\partial p^\lambda}
	\Gamma_{Y\varphi c^\rho|{\substack{p=p_{\rm sym}(-\mu^2) \\ s=1} }}&=& -i\delta^\rho_\lambda\\
	\frac{\partial}{\partial p^\lambda}\Gamma_{{L_\rho}c^\sigma c^\tau|{\substack{p=p_{sym}(-\mu^2) \\ s=1} }}&=&
			-i\kappa(\delta^\rho_\sigma\eta_{\lambda\tau}
				  -\delta^\rho_\tau\eta_{\lambda\sigma}).
\end{eqnarray}
Imposing the $b$-equation of motion (\ref{beq}) still fixes $\alpha_0$ 
and the $b$-amplitude. \\

\section{Invariant differential operators and invariant insertions}
%\P file SymDifo.tex\P\\
Here we develop the concept of BRST-invariant differential operators
and their one-to-one counterparts, BRST-invariant insertions.
One can essentially follow the paper \cite{PS2} and translate from YM to gravity.

Suppose a model satisfies the WI of a linear transformation
\be\label{ltWI}
W^a\Gamma\equiv\int\delta^a\phi\frac{\delta\Gamma}{\delta\phi}=0
\ee
and $\lambda$ is a parameter of the theory (e.g.\ coupling, mass, 
normalization parameter) of which the WI-operator $W^a$ does not
depend. Then $\lambda\partial_\lambda$ commutes with $W^a$, i.e.
\be\label{sc0}
[\lambda\partial_\lambda,W^a]=0.
\ee
Then the action principle tells us that 
\be\label{srt}
\lambda\partial_\lambda\Gamma=\Delta_\lambda\cdot\Gamma .
\ee
Applying $W^a$ to (\ref{srt}) and using (\ref{sc0}) we find
\be\label{sct}
W^a(\Delta_\lambda\cdot\Gamma)= W^a\Delta_\lambda+O(\hbar\Delta)=0,
\ee
which expresses the invariance of $\Delta_\lambda$ under the symmetry 
transformation $W^a$: $\lambda\partial_\lambda$ and $\Delta_\lambda$ are 
called symmetric with respect to the symmetry $W^a$.

For the $\Gamma$-non-linear BRST-symmetry one has to proceed slightly 
differently. We shall call an insertion $\Delta$ BRST-symmetric if to first 
order in $\epsilon$
\begin{eqnarray}
	\mathcal{S} (\Gamma_\epsilon)&=&O(\epsilon^2) \label{scbrst}\\
	{\rm for} \qquad \Gamma_\epsilon&=&\Gamma+\epsilon\Delta\cdot\Gamma
		   \qquad{\rm with}\qquad \mathcal{S} (\Gamma)=0.
\end{eqnarray}
If $\Delta$ is generated by a differential operator $(\ref{srt})$, this
differential operator will be called BRST-symmetric. Writing (\ref{scbrst})
explicitly we have
\be\label{scbrst2}
\ST(\Gamma)+\epsilon \ST_\Gamma\Delta\cdot\Gamma=O(\epsilon^2)
\ee
\be\label{scbrst3}
\ST_\Gamma\equiv\int\left(
 \frac{\delta\Gamma}{\delta K}\frac{\delta}{\delta h}
+\frac{\delta\Gamma}{\delta h}\frac{\delta}{\delta K}
+\frac{\delta\Gamma}{\delta L}\frac{\delta}{\delta c}
+\frac{\delta\Gamma}{\delta c}\frac{\delta}{\delta L}
+\frac{\delta\Gamma}{\delta Y}\frac{\delta}{\delta\varphi}
+\frac{\delta\Gamma}{\delta \varphi}\frac{\delta}{\delta Y}
+b\frac{\delta}{\delta\bar{c}}\right)
+\chi\frac{\partial}{\delta\alpha_0} ,
\ee

i.e.\ the symmetry condition reads
\be\label{scbrst4}
\ST_\Gamma\Delta\cdot\Gamma=0 .
\ee
A comment is in order. Although later we shall exclusively work in Landau gauge, we carry here the gauge parameter $\alpha_0$ along as
preparation for the general solution with arbitrarily many parameters $z_{nk}$. This 
facilitates the formulation of the general version. Actually relevant at the end
are only the formulae with $\alpha_0=\chi=0$.
The explicit form of $\ST_\Gamma$ precisely defines how to perform the 
variation of the fields. \footnote{This formula shows that it is not the
demand ``linearity in $\Gamma$'' which determines its form, but rather
the demand ``correct  transformation of an insertion ''.}
The operator $\ST_\Gamma$ is helpful for rewriting the gauge fixing and 
$\phi\pi$-contributions to the action \eqref{2cmgf}:
\be\label{vfrmgffp}
\Gamma_{\rm gf}+\Gamma_{\phi\pi}
    = \ST_\Gamma\left(-\frac{1}{2\kappa}\int h^{\mu\nu}(x)
              (\partial_\mu\bar{c}_\nu+\partial_\nu\bar{c}_\mu)(y)
	      \Big\{\big( \frac{\Box}{4\pi^2} + m^2 \big)\frac{1}{(x-y)^2}\Big\}
	       -\int\frac{\alpha_0}{2}\eta^{\mu\nu}\bar{c}_\mu b_\nu\right) .
\ee
(Note: the last term creates a contribution which has not been taken into 
account in (\ref{2cmgf}), however in (\ref{varalph}).)
When going over to $Z$, the generating functional for the general Green 
functions,
it is clear, that gauge fixing and $\phi\pi$-term vanish between physical
states, because they are a BRST-variation.\\
A necessary condition for insertions to be BRST-symmetric is obtained
by acting with $\delta/\delta b$ on (\ref{scbrst}): 
\be\label{snn}
G\Delta\cdot\Gamma=S_\Gamma\frac{\delta\Delta\cdot\Gamma}{\delta b},
\qquad G^\rho \equiv\frac{\delta}{\delta\bar{c}_\rho(x)}
+\kappa^{-1}\int dy\,\partial_\lambda\frac{\delta\bar{\Gamma}}{\delta K_{\rho\lambda}(y)}
   \Big\{\big( \frac{\Box}{4\pi^2} + m^2 \big)\frac{1}{(x-y)^2}\Big\} .
\ee
For $b$-independent insertions $\Delta$ one must ensure the homogeneous 
ghost equation
\be\label{ghD}
G\Delta\cdot\Gamma = 0 .
\ee
Using the gauge condition
\be\label{gc}
\frac{\delta\Gamma}{\delta b_\rho}=-\alpha_0 \eta^{\rho\lambda}b_\rho
 +\kappa^{-1}\int dy\,\partial_\mu h^{\mu\rho}(y)\Big\{\big( \frac{\Box}{4\pi^2} + m^2 \big)\frac{1}{(x-y)^2}\Big\} ,
\ee
one can reduce (\ref{snn}) to         
\be\label{rsc}
\mathcal{B}_{\bar{\Gamma}}\Delta\cdot\Gamma=0 .
\ee
In the tree approximation we have called this operator $\brstb$.

Our next task is to construct a {\it basis} for all symmetric insertions 
of dimension 4, $\phi\pi$-charge 0, and independent of $b_\rho$ -- first in the
tree approximation and then to all orders. A systematic way to find them is
to solve the cohomology problem
\be\label{cpr}
\brstb\Delta=0
\ee
for $\Delta$ satisfying
\begin{eqnarray}\label{cpr2}
	\frac{\delta\Delta}{\delta b}=&0,&G\Delta=0\\
	{\rm dim}(\Delta)=&4,&Q_{\phi\pi}(\Delta)=0 .
\end{eqnarray}

Here $\brstb=\mathcal{B}_{\bar{\Gamma}_{\rm class}}$, hence
\begin{eqnarray}\label{lsttrs}%listoftransformations
	  \brstb&=& \brsts \qquad{\rm on\, all\, elementary\, fields}\\
\brstb H_{\mu\nu}&=&\frac{\delta\bar{\Gamma}_{\rm cl}}{\delta h^{\mu\nu}}
	     =\frac{\delta\Gamma^{\rm class}_{\rm inv}}{\delta h^{\mu\nu}}
	     -\kappa(H_{\lambda\mu}\partial_\nu c^\lambda
	     +H_{\lambda\nu}\partial_\mu c^\lambda
	     +\partial_\lambda(H_{\mu\nu}c^\lambda))\label{lsttrsH}\\
	\brstb Y&=&\frac{\delta\bar{\Gamma}_{\rm cl}}{\delta\varphi}=
         \frac{\delta\Gamma^{\rm class}_{\rm inv}}{\delta \varphi}
	-\partial_\lambda Yc^\lambda
	        -\frac{3}{4}Y\partial_\lambda c^\lambda    \\
\brstb L_\rho&=&\frac{\delta{\bar\Gamma}_{\rm cl}}{\delta c^\rho}=
	    \kappa(2\partial^\lambda H_{\lambda\rho}
	    +2\partial_{\lambda'}(H_{\rho\lambda}h^{\lambda'\lambda}
	    +H_{\lambda'\lambda}\partial_\rho h^{\lambda\lambda'}))\\
	   &&\qquad\quad  -\kappa(L_\lambda\partial_\rho c^\lambda
		    +\partial_\lambda(L_\rho c^\lambda))  \label{lsttrsL}\\
	   &&\qquad\quad+Y\partial_\rho\varphi
	-\frac{1}{4}\partial_\rho(Y\varphi ) \label{lsttrsY}. 
\end{eqnarray}
In order to proceed we first separate the $\alpha_0$-dependence
\be\label{s0d}
\Delta=\chi_0\Delta_- +\Delta_0 .
\ee
We now define
\be\label{bbr}
\bar{\brstb}=\left\{\begin{array}{l}
	      \brstb\qquad{\rm on}\quad h,c,H,L,Y\\
              0\qquad{\rm on}\quad \alpha_0
              \end{array}
       \right.
\ee
and note that
\be\label{bbra}
\partial_{\alpha_0}(\brstb\psi)=0\qquad{\rm for}\quad \psi=h,c,H,L,Y
\ee
with $\bar{\brstb}^2$=0, since $\bar{\Gamma}_{\rm cl}$ is independent of $\alpha_0$.
(\ref{cpr}) implies
\be\label{spr}
\bar{\brstb}\Delta_- -\partial_{\alpha_0}\Delta_0=0 \qquad \bar{\brstb}\Delta_0=0,
\ee
hence
\be\label{spr2}
\Delta=\brstb\hat{\Delta}_- +\hat{\Delta}_0.
\ee
Here $\hat{\Delta}_0$ is $\alpha_0$-independent and $\bar{\brstb}$-invariant.
Since $\bar{c}$ does not occur, a negative $\phi\pi$-charge can only be
generated by external fields, hence  
\be\label{lfe}
\hat{\Delta}_- = \int(f_H(\alpha_0)H_{\mu\nu}h^{\mu\nu}
      +f_Y(\alpha_0)Y\varphi+f_L(\alpha_0)L_\rho c^\rho)
\ee
which is the precise analogue of (4.19) in \cite{PS23},
is certainly a solution. However in the present case the field $h^{\mu\nu}$
has canonical dimension zero, whereas its counterpart in Yang-Mills theory, 
the vector field $A_\mu$ has dimension one. So every function 
$\mathcal{F}^{\mu\nu}(h)$ is also a solution. For the time being we continue
with (\ref{lfe}) and refer for the discussion of the general solution to papI.\\
It is worth solving the subproblem
\be\label{spr3}
\partial_{\alpha_0}\hat{\Delta}_0=0 \qquad \bar{\brstb}\hat{\Delta}_0=0
\ee
explicitly. 
We start listing the contributions to $\hat{\Delta}_0$ ordered by their
external field dependence, i.e.  
\be\label{Yd}
\hat{\Delta}_0=f_Y(0)\int Yc^\lambda\varphi+\cdots (\rm indep.\, of\, Y), 
\ee
where $f_Y(0)$ is an arbitrary number independent of $\alpha_0$. With 
(\ref{lsttrsY})
this can be rewritten as 
\be\label{Yd1}
\hat{\Delta}_0=f_Y(0)\bar{\brstb}(\int Y\varphi)+\cdots ({\rm indep.\,of\,Y})
\ee
or as
\be\label{Yd2}
\hat{\Delta}_0=\brstb\int(f_Y(0))Y\varphi)+\cdots ({\rm indep.\,of\,Y})
\ee
It is to be noted that to the $\brstb$-variation of $Y$ the 
invariant kinetic term for $\varphi$ contributes, hence will turn ot to be
a variation.\\
The analogous procedure for the $L$ terms leads to
\be\label{Ld}
\hat{\Delta}_0=-f_L(0)\kappa\int L_\rho c^\lambda\partial_\lambda c^\rho
			 +\cdots({\rm indep.\, of}\, L),
\ee
where $f_L(0)$ is an arbitrary number independent of $\alpha_0$. 
With (\ref{lsttrsL}) this term can be rewritten as
\be\label{Ld2}
\hat{\Delta}_0=f_L(0)\bar{\brstb}(\int L_\rho c^\rho)
                                +\cdots({\rm indep.\, of}\, L)
\ee
\be\label{Ld3}
\hat{\Delta}_0= \brstb \int(f_L(0)L_\rho c^\rho)
                                +\cdots({\rm indep.\, of}\, L) .
\ee
We next make explicit the $H$-dependence
\be\label{Hd}
\hat{\Delta}_0= \brstb \int(f_L(0) L_\rho c^\rho)
	  +\int H_{\mu\nu}F_{(+)}^{\mu\nu}(h,c)+\cdots(L,H)-{\rm indep.}
\ee
The postulate (\ref{spr3}) reads
\begin{eqnarray}\label{Hd2}
	0&=&\bar{\brstb}\hat{\Delta}_0=
       \int\Big(\frac{\delta\bar{\Gamma}_{\rm cl}}{\delta h}F^{(+)}
-H\bar{\brstb}F^{(+)}\Big)+(L,H)-{\rm indep.}\\
	&=:&-\int H\mathcal{C}F^{(+)}+(L,H)-{\rm indep.}
\end{eqnarray}
and defines a transformation $\mathcal{C}$ as the coefficient of $H$ in
(\ref{Hd}):
\be\label{mthcC}
\mathcal{C}F_{(+)}=\bar{\brstb}F^{(+)}
      +\kappa(\partial_\lambda c^\mu F^{\nu\lambda}_{(+)}
             +\partial_\lambda c^\nu F^{\mu\lambda}_{(+)}
	     -c^\lambda\partial_\lambda F^{\mu\nu}_{(+)}).
\ee
This transformation is nilpotent and satisfies, due to (\ref{Hd2}),
\be\label{mthcC2}
\mathcal{C}F_{(+)}=0
\ee
One solution is 
\be\label{sF}
F^{\mu\nu}_{(+)}=\mathcal{C}(f_H(0)h^{\mu\nu}).
\ee
Since
\be\label{sF2}
\mathcal{C}(h^{\mu\nu})=
              \kappa(-\partial^\mu c^\nu-\partial^\nu c^\mu) ,
\ee
it fits correctly to the $H$-dependent part of $(\ref{lsttrsH})$ in
(\ref{Hd2}).
One thus arrives for this solution at
\be\label{sHd3}
\bar{\brstb}\int f_H(0)H_{\mu\nu}h^{\mu\nu}
		       =\int H_{\mu\nu}\mathcal{C}(f_H(0)h^{\mu\nu}) ,
\ee
i.e.\ the $H$-dependent part in $\hat{\Delta}_0$ is also a variation.
As mentioned above this is not the most general solution, but the outcome
is analogous. The details can be found in papI.\\

The remaining contributions to $\hat{\Delta}_0$ depend only on $h$ and must not depend on $\alpha_0$. 
The only invariants are those terms appearing in 
$\Gamma^{\rm class}_{\rm inv}$ which come with the couplings: 
$c_1,c_2,c_3\kappa^{-2},c_R,\lambda$. 
They are not variations, but constitute obstruction terms to the $\bar{\brstb}$-cohomology. 
Altogether we thus have  
\begin{eqnarray}\label{slcm0}
\Delta_0&=& \brstb \int\Bigl(f_Y(0)Y\varphi+f_L(0)L_\rho c^\rho
		       +f_H(0)H_{\mu\nu}h^{\mu\nu}\Bigr)\\
	&&+\int\,\lbrace(\sqrt{-g}(\hat{c}_3\kappa^{-2}R
	+\hat{c}_1R^{\mu\nu}R_{\mu\nu}+\hat{c}_2R^2)) 
	    +(-g)^{1/4}\frac{\hat{c}_R}{2}\varphi^2 R 
	    -\frac{\lambda}{4!}\varphi^4\rbrace
\end{eqnarray}
(The factors $\hat{c}$ are independent of $\alpha_0$.)
In tree approximation we end up with eight  invariant insertions of 
dimension 4
and $\phi\pi$-charge 0, which are independent of $b_\rho$ and satisfy the 
ghost equation:  
\begin{eqnarray}\label{sns5}
    \Delta'_Y&=&\brstb\left(f_Y(\alpha_0)\int Y\varphi\right)\\
	\Delta'_L&=&\brstb\left(f_L(\alpha_0)\int L_\rho c^\rho\right)\\
	\Delta'_H&=&\brstb\left(f_H(\alpha_0)\int H_{\mu\nu} h^{\mu\nu}\right)\\
	\Delta_{c_3}&=&c_3\kappa^{-2}\int\sqrt{-g}\kappa^{-2}R
  \quad\,	\Delta_{c_1}=c_1\int\sqrt{-g}R^{\mu\nu}R_{\mu\nu}
  \quad\, \Delta_{c_2}=c_2\int\sqrt{-g}R^2 \\
	     \Delta_R&=&\frac{c_R}{2}\int\,(-g)^{1/4}\varphi^2R \quad\,
\Delta_\varphi=\frac{\lambda}{4!}\int\,\varphi^4
\end{eqnarray}
(Here we 
renamed the couplings of the non-variations.)\\
In higher orders we may define easily invariant insertions for those which 
come with the couplings:  
\be\label{hghcpls}
\Delta_{c_i}:=c_i\frac{\partial}{\partial c_i}\Gamma\quad 
                                   (i=1,2,3,R,\lambda \quad{\rm no\,\, sum}),
\ee
however it is clear that the $(s-1)$-dependent normal products
$c_{31}[\kappa^{-1}m\int\sqrt{-g}R\,]^4_4$ and 
$c_{32}1/2[m^2\int\sqrt{-g}R\,]^4_4$
also belong to the basis in higher orders and make part of $\Gamma_{\rm eff}$.
Hence we define them also as invariant by the respective derivation with
respect to their coupling
\be\label{gnbss}%generalbasis
\Delta_{c_{31}}:=c_{31}\frac{\partial}{\partial c_{31}}\Gamma \qquad
\Delta_{c_{32}}:=c_{32}\frac{\partial}{\partial c_{32}}\Gamma .
\ee
Accordingly we change the notation $c_3\rightarrow c_{30}$.
The other terms we also try to represent as symmetric {\sl differential}
operators acting on $\Gamma$.\\
We rewrite $\Delta'_L$:
\begin{eqnarray}\label{Ldo}
	\Delta'_L= \brstb\left(f_L(\alpha_0)(\alpha_0)\int L_\rho c^\rho\right)
  &=&\chi f'_L\int Lc
     +f_L\int\left(\frac{\delta\bar{\Gamma}_{\rm cl}}{\delta c}
     +L\frac{\delta\bar{\Gamma}_{\rm cl}}{\delta L}\right)\\
  &=&\chi f'_L\int Lc
     +f_L\int\left(-c\frac{\delta\bar{\Gamma}_{\rm cl}}{\delta c}
     +L\frac{\delta\bar{\Gamma}_{\rm cl}}{\delta L}\right)\\
	&=&-f_L\mathcal{N}_L\Gamma_{\rm cl}
     +\chi f'_L\int Lc,
\end{eqnarray}
where $\mathcal{N}$ denote a leg-counting operator. This suggests defining
$\Delta_L$ to all orders by
\begin{eqnarray}\label{Ldoh}
\Delta_L\cdot\Gamma&=&f_L(\alpha_0)\mathcal{N}_L\Gamma-\chi_0 f'_L\int Lc,\\
\mathcal{N}_L&\equiv&\int\Big(c\frac{\delta}{\delta c}-L\frac{\delta}{\delta L}\Big)
    =N_c-N_L .
\end{eqnarray}
It is to be noted that the $\chi$-dependent term in (\ref{Ldoh}) is well 
defined since $L$ is an external field, hence the expression is linear in the
quantized field (c). $\Delta_L$ does obviously not depend on $b_\rho$, it
satisfies the ghost equation and it fulfills (\ref{rsc}), since it can be 
written as
\be\label{fdL}
\Delta_L\cdot\Gamma=-\mathcal{B}_{\bar{\Gamma}}\left(f_L\int Lc\right),
\ee
and since $\mathcal{B}_{\bar{\Gamma}}$ is nilpotent. Hence it is a 
BRST-symmetric operator to all orders.\\
Analogously  
\begin{align}\label{fdY}
\Delta_Y\cdot\Gamma
	=&\mathcal{B}_{\bar{\Gamma}}\left(f_Y(\alpha_0)\int Y\varphi\right)=
f_Y(\alpha_0)\mathcal{N}_Y\Gamma-\chi_0 f'_Y\int Yc\\
	\mathcal{N}_Y\equiv&\int\Bigl(\varphi\frac{\delta}{\delta \varphi}
	-Y\frac{\delta}{\delta Y}\Bigr)=N_\varphi-N_Y
\end{align}

Finally we have to extend $\Delta_H'$. We first rewrite it in the form
\be\label{Hdo}
\Delta_H'= \brstb\left(f_H(\alpha_0)\int H_{\mu\nu}h^{\mu\nu}\right)
	=f_H N_H\bar{\Gamma}_{\rm cl}-f_HN_H\Gamma_{\rm cl}
		+\chi f'_H\int H_{\mu\nu}h^{\mu\nu}.
\ee
Next we go over to $\Gamma_{\rm cl}$ in the variables $K$ and $\bar{c}$:
\begin{eqnarray}\label{Hon}
	\Delta_H'&=&f_H(N_h-N_K-N_b-N_{\bar{c}}
	+2\alpha_0\partial_{\alpha_0}+2\chi\partial_\chi)\Gamma_{\rm cl}\\
	&&+\chi f'_H\Big(\int \Big( Kh-\bar{c}\frac{\delta\Gamma_{\rm cl}}{\delta b}\Big)
	   +2\alpha_0\frac{\partial}{\partial \chi}\Gamma_{\rm cl}\Big) .
\end{eqnarray}
This suggests as definition of $\Delta_H$ to all orders
\begin{align}\label{DltaH}%Delta H
	\Delta_H\cdot\Gamma:=&f_H\mathcal{N}_K\Gamma\\
	  &+\chi f'_H\Bigl(\int\,\Bigl( Kh-\bar{c}
	     \frac{\delta\Gamma_{\rm cl}}{\delta b}\Bigr) 
              +2\alpha_0\frac{\partial}{\partial\chi}\Gamma\Bigr),\\
	\mathcal{N}_K\equiv&N_h-N_K-N_b-N_{\bar{c}}
+2\alpha_0\partial_{\alpha_0} +2\chi\partial_\chi
\end{align}
Or else
\be\label{sDltaH}
\Delta_H\cdot\Gamma:=\mathcal{S}_\Gamma\Bigl(f_H(\alpha_0)      
	       \Bigl(\int\Bigl(Kh-\bar{c}\frac{\delta\Gamma}{\delta b}\Bigr)
	     +2\alpha_0\frac{\partial\Gamma}{\partial\chi}\Bigr)\Bigr)
\ee
In view of 
\be\label{cstcy}
\mathcal{S}_\Gamma\mathcal{S}_\Gamma=0
\ee
for all $\Gamma$ with $\mathcal{S}(\Gamma)=0$, $\Delta_H$ is BRST symmetric 
once we have verified that it is independent of $b_\rho$ and satisfies the
ghost equation.
\be\label{bndpnt}
\frac{\delta}{\delta b}(\Delta_H\cdot\Gamma)=0
\ee
is readily checked in the form (\ref{DltaH}).
\be\label{ghchk}
G(\Delta_H\cdot\Gamma)=0
\ee
is best checked in the form (\ref{sDltaH}) by observing that 
\be\label{2ghchk}
G\Bigl(\int\Bigl(Kh-\bar{c}\frac{\delta\Gamma}{\delta b}\Bigr)
                   +2\alpha_0\frac{\partial\Gamma}{\partial\chi}\Bigr)=0, 
\ee	
and
\be\label{ghsg}
\{G,\ST_\Gamma\}=0
\ee
(this latter property being due to $G\Gamma=-1/2 \chi b$).

To summarize in compact notation we denote the above symmetric differential
operators by
\be\label{tns}%notations
\nabla_i\in \{c_1\partial_{c_1}, c_2\partial_{c_2}, 
c_{30}\partial_{c_{30}}, c_{31}\partial_{c_{31}}, c_{32}\partial_{c_{32}}, 
c_{\rm R}\partial_{c_{\rm R}},\lambda\partial_\lambda,
             \mathcal{N}_H, \mathcal{N}_L, \mathcal{N}_Y \}
\ee
and have with (\ref{hghcpls}),(\ref{gnbss}), (\ref{Ldoh}), (\ref{fdY}) and 
(\ref{DltaH})
defined a basis of symmetric insertions to all orders by
\be\label{gnp}
\nabla_i\Gamma \stackrel{.}{=} \Delta_i\cdot\Gamma.
\ee
In parentheses we note that the kinetic term of the scalar field $\varphi$
does not explicitly appear in this basis. It is taken care of via 
$\mathcal{N}_Y$ and the respective normalization condition for $Yc\varphi$.\\
The fact that symmetric differential operators and symmetric insertions
are in one-to-one correspondence just means that adding symmetric 
counterterms $\Delta_i$ to $\Gamma$ is renormalizing the corresponding
quantity $i$ indicated by $\nabla_i$ of the theory. Fixing the arbitrary 
parameters in the symmetric insertions (\ref{sns5}) is again performed by
satisfying normalization conditions and the present analysis shows
that the conditions (\ref{highnorm}) are appropriate. In higher orders
the Euclidean point $-\mu^2$ is relevant. Once one has satisfied these 
normalization conditions the theory is completely fixed.
$\alpha_0=0$ and $\chi=0$ will be chosen at the end (Landau gauge).\\

\section{Removing auxiliary mass dependence via Zimmermann Identities}
%\P file auxmass.tex \P\\
Above we have introduced amongst the symmetric insertions several which depend
on the auxiliary mass. Here we study to which extent they can be effectively 
removed by using ZI's.

\subsection{Shift} 
In (\ref{ivce}) we replaced
$c_3\kappa^{-2}
{\rm within}\quad\gamma^{(r)}_{KL},\, r=2,K=L=T$ by 
$c_3\kappa^{-2} \rightarrow c_{30}\kappa^{-2}+m\kappa^{-1}c_{31}
+\frac{1}{2}m^2 c_{32},
$
where $m\equiv M(s-1)$.
On the level of symmetric insertions this replacement corresponds to enlarging the basis of naively BRST-invariant insertions with $\rho=\delta=4$  
by $c_{31}m\kappa^{-1}\int\sqrt{-g}R$ and
    $c_{32}\frac{1}{2} m^2\int\sqrt{-g}R$, which are to be taken into account in 
$\Gamma_{\rm eff}$ .\\
Then the question is, whether one can via ZI's eliminate the $m$-terms and
maintain invariance. The sought invariant $[...]^4_4$ insertions are 
defined to all orders as symmetric insertions via the invariant derivatives
\begin{eqnarray}\label{smns}
\big[\kappa^{-2}\int\sqrt{-g}R\,\big]^{4}_{4}&=&
	                 \frac{\partial}{\partial c_{30}}\Gamma\\
\big[\kappa^{-1}\int\sqrt{-g}m R\,\big]^{4}_{4}&=&
	                 \frac{\partial}{\partial c_{31}}\Gamma\\
\big[\int\sqrt{-g}\frac{1}{2}m^2R\,\big]^4_4&=&
	                 \frac{\partial}{\partial c_{32}}\Gamma
\end{eqnarray}
and the symmetric counting operators $\mathcal{N}_{Y,H,L}$.
The relevant ZI's have the form  
\begin{eqnarray}\label{sZ}
\big[\kappa^{-2}\int\sqrt{-g}R\,\big]^{4}_{4}&=&
	\big[\kappa^{-2}\int\sqrt{-g}R\,\big]^{3}_{3}+[...]^4_4\label{sZ0}\\
	{\rm with}\quad [...]^4_4&=&[\int\,(\sqrt{-g}(\kappa^{-2}u_{0}R
  +u_{31}m\kappa^{-1}R+u_{32}\frac{1}{2}m^2R
	 + u_1 R^{\mu\nu}R_{\mu\nu}+u_2R^2)\nonumber\\
	&&+(-g)^{1/4}u_R \varphi^2R +u_\varphi\varphi^4)\\
	&&+u_\varphi\,\mathcal{N}_Y\,+u_h\mathcal{N}_H+u_c\,\mathcal{N}_L]^4_4\\
\big[\kappa^{-1}\int\sqrt{-g}mR\,\big]^{4}_{4}&=&
m\big[\kappa^{-1}\int\sqrt{-g}\kappa^{-1}R\,\big]^{3}_{3}+[...]^4_4\label{sZ1}\\
	{\rm with}\quad [...]^4_4&=&[\int\,(\sqrt{-g}(\kappa^{-2}v_{30}R
	+v_0m\kappa^{-1}R+v_{31}\frac{1}{2}m^2R+v_1 R^{\mu\nu}R_{\mu\nu}+v_2R^2)\\
       &&+(-g)^{1/4}v_R\varphi^2R +v_\varphi\varphi^4)\\
      &&+v_\varphi\,\mathcal{N}_Y+v_h\,\mathcal{N}_H+v_c\,\mathcal{N}_L]^4_4
\end{eqnarray}
and  

%\newpage
\begin{eqnarray}	                
	\big[\int\sqrt{-g}\frac{1}{2}m^2 R\,\big]^4_4&=&
     m\big[\int\sqrt{-g}\frac{1}{2}mR\,\big]^3_3+[...]^4_4
	                                                \label{sZ2}\\
	{\rm with}\quad [...]^4_4&=&[\int\,(\sqrt{-g}(\kappa^{-2}w_{30}R
	+w_{31}m\kappa^{-1}R+w_0\frac{1}{2}m^2R\nonumber
        + w_1 R^{\mu\nu}R_{\mu\nu}+w_2R^2) \\
	&&+(-g)^{1/4}w_R\varphi^2R +w_\varphi\varphi^4)\\
	&& +w_\varphi\,\mathcal{N}_Y +w_h\,\mathcal{N}_H+w_c\,\mathcal{N}_L]^4_4 .
\end{eqnarray}
All coefficients $u,v,w$ are of order $\hbar$. The terms multiplied by 
$u_0,v_0,w_0$ resp.\ will be absorbed on the resp.\ lhs
and then the resp.\ line  divided by $1-u_0,1-v_0,1-w_0$, such that the
normal products on the rhs have the factors 
      $(1-u_0)^{-1}, (1-v_0)^{-1}, (1-w_0)^{-1}$
in the resp.\ line. From this representation it is then obvious that all 
$[...]^3_3$ insertions on the rhs are symmetric, because all other insertions
are symmetric. Since the relevant determinant in this linear system of 
equations is clearly non-vanishing, one can solve for all hard insertions
$[\int\sqrt{-g}R(\kappa^{-2}, m\kappa^{-1},\frac{1}{2}m^2)]^4_4$ in terms of 
the soft one's together with $(c_1,c_2,c_R,c_\varphi,\mathcal{N}_{Y,H,L})$-terms. But those soft insertions
which contain the factor $m$ vanish at $s=1$, hence all hard $m$-dependent
insertions have been eliminated. And the hard insertion 
$[\kappa^{-2}\int\sqrt{-g}R]^4_4$ has been effectively replaced by its soft 
counterpart.
These considerations are crucial for deriving the parametric
differential equations in symmetric form and without dependence from
the auxiliary mass $M$ at $s=1$.

\subsection{Push}
Next we consider the problem of removing Push by using appropriate ZI's. 
First we treat the contributions of Push to $\Gamma^{\rm class}_{\rm inv}$ (cf. (\ref{mcoffs})).
They occur in the second power of $h$ and have the form (see (\ref{gmps1})),(\ref{gmps2}))
\be\label{pshnv}
\Gamma_{(hh)}(m^2)=
\int\,h^{\mu\nu}(m^2\hat{\gamma}^{(2)}_{\rm TT}P^{(2)}_{\rm TT}
           +m^2\hat{\gamma}^{(0)}_{\rm TT}P^{(0)}_{\rm TT})_{\mu\nu\rho\sigma}
                                h^{\rho\sigma}.
\ee
In higher orders we have just the same terms, but now to be understood as
normal products $[...]^4_4$ in $\Gamma_{\rm eff}$. We use the ZI
\begin{multline}\label{pZi}
	[\int\,h^{\mu\nu}(m^2\hat{\gamma}^{(2)}_{\rm TT}P^{(2)}_{\rm TT}
           +m^2\hat{\gamma}^{(0)}_{\rm TT}P^{(0)}_{\rm TT})_{\mu\nu\rho\sigma}
				h^{\rho\sigma}]^4_4\cdot\Gamma_{(hh)}
	\phantom{M(s-1)}			\\
	=M(s-1)[\int\,h^{\mu\nu}(m\hat{\gamma}^{(2)}_{\rm TT}P^{(2)}_{\rm TT}
           +m\hat{\gamma}^{(0)}_{\rm TT}P^{(0)}_{\rm TT})_{\mu\nu\rho\sigma}
		h^{\rho\sigma}]^3_3\cdot\Gamma_{(hh)}
	      +[{\rm corr.s}]^4_4\cdot\Gamma_{(hh)}.
\end{multline}
Here the $\hat{\gamma}$'s are interpreted as differential operators and
$m\equiv M(s-1)$ is to be recalled.
The corrections comprise first of all the starting term from the lhs with
a coefficient $q=O(\hbar)$. We bring it to the lhs and divide by $1-q$.
This yields
%\newpage
\begin{multline}\label{pZi2}
	[\int\,h^{\mu\nu}(m^2\hat{\gamma}^{(2)}_{\rm TT}P^{(2)}_{\rm TT}
           +m^2\hat{\gamma}^{(0)}_{\rm TT}P^{(0)}_{\rm TT})_{\mu\nu\rho\sigma}
				h^{\rho\sigma}]^4_4\cdot\Gamma_{(hh)}
	\phantom{M(s-1)}\\
           = \frac{M(s-1)}{1-q}
 [\int\,h^{\mu\nu}(m\hat{\gamma}^{(2)}_{\rm TT}P^{(2)}_{\rm TT}
           +m\hat{\gamma}^{(0)}_{\rm TT}P^{(0)}_{\rm TT})_{\mu\nu\rho\sigma}
		h^{\rho\sigma}]^3_3\cdot\Gamma_{(hh)}
	      +\frac{1}{1-q}[{\rm corr.s}]^4_4\cdot\Gamma_{(hh)}.
\end{multline}
As correction terms appear the $hh$-vertex functions with all 
$[...]^4_4$-insertions. We now can demand $\brsts_0$-invariance because this
is a linear transformation. Amongst the $\hat{\gamma}^{(r)}_{\rm K,L}$-
contributions precisely those with $r=2,0; K=L=T$ are $\brsts_0$-invariant,
hence they have been absorbed already. The other contributions go with the
symmetric differential operators $\mathcal{N}_{\rm Y,H,L}$. These are however
BRST-variations and thus vanish between physical states. Therefore this
part of Push does at $s=1$ not contribute to physical quantities. 

The second (and last) appearance of Push is within gauge fixing and 
$\phi\pi$-terms. 
\begin{eqnarray}\label{pshgf}
(\Gamma_{\rm gf}+\Gamma_{\phi\pi})(m^2))&=&
	  -\frac{1}{2}\int \Big( \frac{1}{\kappa}h^{\mu\nu}(x)
    (\partial_\mu b_\nu+\partial_\nu b_\mu)(y)\frac{m^2}{(x-y)^2}\nonumber\\
&&\qquad+ D^{\mu\nu}_\rho c^\rho(x)(\partial_\mu\bar{c}_\nu
	 +\partial_\nu\bar{c}_\mu)(y)\frac{m^2}{(x-y)^2} \Big)\nonumber\\
&=&-\frac{1}{2}\int\, \brsts_\Gamma \Big(h^{\mu\nu}(x)(\partial_\mu\bar{c}_\nu
			     +\partial_\nu\bar{c}_\mu)(y)\frac{m^2}{(x-y)^2} \Big).
\end{eqnarray}
The product in the last line is point split in $(x\leftrightarrow y)$. 
Divergences can be 
developed at coinciding points in such a way that they can be controlled by a ZI
\be\label{dcs}%divergencecontrosplit
[h^{\mu\nu}(x)(\partial_\mu\bar{c}_\nu
		       +\partial_\nu\bar{c}_\mu)(y)m^2]^4_4\cdot\Gamma=
m[h^{\mu\nu}(x)(\partial_\mu\bar{c}_\nu
               +\partial_\nu\bar{c}_\mu)(y)m]^3_3\cdot\Gamma
	+\,[{\rm corr.s}]^4_4\cdot\Gamma
\ee
Amongst the corrections, again, appears the normal product of the lhs, which can 
be absorbed there, such that on the rhs only all other insertions
of dimension 4 and $\phi\pi$-charge $-1$ show up. These are 
$Y\varphi,K_{\mu\nu}h^{\mu\nu},L_\rho c^\rho$ which are all naively defined because
they are linear in the quantized fields. At $s=1$ they are the only surviving
terms which contribute in (\ref{pshgf}) and then eventually vanish after
integration between physical states.\\

\subsection{Auxiliary mass of the scalar field}
Since the respective term is not $\brsts$ invariant, it has to be treated
separately. Relevant is the Zimmermann identity
\be\label{zdscf}
\lbrack\int M^2(s-1)^2\varphi^2\rbrack^4_4= 
\frac{M(s-1)}{1+q'}\lbrack\int M(s-1)\varphi^2\rbrack^3_3
                               +{\rm corr.s}
\ee
Here the correction terms consist of all $\lbrack\cdot\cdot\cdot\rbrack^4_4$
which form the respective basis. These are first of all the symmetric terms
$\int(\sqrt{-g}(R^{\mu\nu}R_{\mu\nu},
R^2), \int(-g)^{1/4}\varphi^2R,\int \frac{1}{4!}\varphi^4$ and the differential
operators $\mathcal{N}_{H,L,Y}$; however also all non-symmetric counterterms
which contribute to $\Gamma_{\rm eff}$. They all come with coefficients 
of order $O(\hbar)$.\\

\section{The invariant parametric differential equations}
%\P file InvDifEqs\P\\ 
\subsection{The Lowenstein-Zimmermann equation}
Green functions must be independent of the auxiliary mass $M$ at $s=1$, so
one has to know the action of $M\partial_M$ on them. Since the ST-identity
does not depend on $M$, $M\partial_M$ is a BRST-invariant differential 
operator and can be expanded in the basis provided by $(\ref{tns})$.
In fact with the ZI's (\ref{sZ1}) and (\ref{sZ2}) and the discussion
there we can consider the basis of symmetric differential operators to
be given by $c_{30} \partial_{c_{30}}, c_1 \partial_{c_1}, c_2\partial_{c_2},c_R\partial_{c_R},\lambda\partial_\lambda$ complemented
with the symmetric counting operators $\mathcal{N}_{Y,H,L}$.
Furthermore we have shown that the contributions
coming from Push (\ref{pZi2}) and the contributions from Shift
go at most into the symmetric counting operators. Hence  
\be\label{sMdM}
M\partial_M\Gamma=(-\beta^{\rm LZ}_{30} c_{30} \partial_{c_{30}}
-\beta^{\rm LZ}_{1} c_1 \partial_{c_{1}}
-\beta^{\rm LZ}_{2} c_2 \partial_{c_{2}}
-\beta^{\rm LZ}_{c_R} \partial_{c_{R}}
-\beta^{\rm LZ}_\lambda \partial_\lambda
   +\gamma^{\rm LZ}_Y\mathcal{N}_Y+\gamma^{\rm LZ}_h\mathcal{N}_H+\gamma^{\rm LZ}_c\mathcal{N}_L)\Gamma .
\ee
The coefficient functions $\beta^{\rm LZ},\gamma^{\rm LZ}$ can be determined by
testing on the normalization conditions. 
The test on \eqref{sMdM} involving external fields  
\be\label{Ycc1}
\frac{\partial}{\partial p^\lambda}\Gamma_{Y c^\rho c^\sigma}
      \,|_{\substack{ p=p_{\rm sym}(-\mu^2) \\ s=1} }
      =-i\eta_{\rho\sigma}
 \ee
implies  
\be\label{Yccp}
M\partial_M\,\partial_p \Gamma_{Ycc}\,|_{\substack{ p=p_{\rm sym}(-\mu^2) \\ s=1}}
-\gamma^{\rm LZ}_\varphi(\partial_p \Gamma_{Ycc}\,|_{\substack{ p=p_{\rm sym}(-\mu^2) \\ s=1}})=0 .
\ee
Since the $M$-derivative in the first term is not in conflict with going to the 
argument of $\Gamma$, the first term vanishes and hence 
$\gamma^{\rm LZ}_\varphi=0$.

The test on
\be\label{Lcc1}
\frac{\partial}{\partial p^\lambda}\Gamma_{L_\rho c^\sigma c^\tau}
	       \,|_{\substack{ p=p_{\rm sym}(-\mu^2) \\ s=1} }	
      =-i\kappa(\delta^\rho_\sigma\eta_{\lambda\tau}
               -\delta^\rho_\tau\eta_{\lambda\sigma})
\ee
implies
\be\label{Lccp}%Lccperformed
M\partial_M\,\partial_p \Gamma_{Lcc}\,|_{\substack{ p=p_{\rm sym}(-\mu^2) \\ s=1}}
-\gamma^{\rm LZ}_c(\partial_p \Gamma_{Lcc}\,|_{\substack{ p=p_{\rm sym}(-\mu^2) \\ s=1}})=0 .
\ee
Since the $M$-derivative in the first term is not in conflict with going
to the argument of $\Gamma$ the first term vanishes and hence
$\gamma^{\rm LZ}_c=0$.
Quite analogously we may proceed for
\be\label{Kc1}
\frac{\partial}{\partial p^\sigma}\Gamma_{K^{\mu\nu}c_\rho}
	                              \,|_{\substack{ p^2=-\mu^2 \\ s=1}}
      =-i\kappa(\eta^{\mu\sigma}\delta_\rho^\nu+\eta^{\nu\sigma}\delta_\rho^\mu
                              -\eta^{\mu\nu}\delta_\rho^\sigma).
\ee
Here this test on (\ref{sMdM}) yields
\be\label{Kcp}%Kcperformed
M\partial_M\,\partial_p \Gamma_{Kc}\,|_{\substack{ p^2=-\mu^2 \\ s=1}}
-\gamma^{\rm LZ}_h (-\partial_p \Gamma_{Kc}\,|_{\substack{ p^2=-\mu^2 \\ s=1}})
-\gamma^{\rm LZ}_c (-\partial_p \Gamma_{Kc}\,|_{\substack{ p^2=-\mu^2 \\ s=1}})=0 .
\ee
With the same argument as before, $\gamma^{\rm LZ}_c=0$ and 
$\gamma^{\rm LZ}_h=0$ follows.\\

The next test reads
\be\label{Ycphi1}
\frac{\partial}{\partial p^\sigma}\Gamma_{Yc_\rho\varphi}
				   \,|_{\substack{ p=p_{\rm sym}(-\mu^2) \\ s=1}}
      =-i\delta_\rho^\sigma.
\ee			      
\be\label{Ycphi}%Ycphi performed
M\partial_M\,\partial_p \Gamma_{Yc\varphi}\,|_{\substack{ p=p_{\rm sym}(-\mu^2)  \\ s=1}}
-\gamma^{\rm LZ}_h (-\partial_p \Gamma_{Yc\varphi}\,|_{\substack{ p=p_{\rm sym}(-\mu^2) \\ s=1}})
-\gamma^{\rm LZ}_c (-\partial_p \Gamma_{Yc\varphi}\,|_{\substack{ p=p_{\rm sym}(-\mu^2) \\ s=1}})=0 .
\ee
With the same argument as before, $\gamma^{\rm LZ}_\varphi=0$ follows.\\ 

For obtaining the $\beta$-functions we use the normalization conditions
(\ref{highnorm}) for $\gamma^{(2)}_{\rm TT}$ and $\gamma^{(0)}_{\rm TT}$,
for $\beta_{c{_R}}$ and $\beta_\lambda$ (\ref{highnorm1}). 
The test
\be\label{hTT3}
\frac{\partial}{\partial p^2}\gamma^{(2)}_{\rm TT}\,|_{\substack{ p^2=0 \\ s=1}}
=c_{30}\kappa^{-2}
\ee
implies
\be\label{hTT3p}
M\partial_M\frac{\partial}{\partial p^2}\gamma^{(2)}_{\rm TT}\,|_{\substack{ p^2=0 \\ s=1}}
+c_{30}\kappa^{-2}\beta^{\rm LZ}_{c_{30}}=0 .
\ee
Since the normalization does not involve $M$, the first term is zero, hence 
$\beta^{\rm LZ}_{c_{30}}=0$.

Tests on the normalization conditions (\ref{highnorm1}) which fix $c_R, \lambda$
\begin{eqnarray}\label{normr}
\frac{\partial^2}{\partial p^\rho\partial p^\sigma}
	    \Gamma_{\varphi\varphi h^{\mu\nu}}\,|_{\substack{ p=p_{\rm sym}(-\mu^2) \\ s=1}}
            &=&\frac{c_R}{2}(\delta^\mu_\rho\delta^\nu_\sigma+\delta^\mu_\sigma\delta^\nu_\rho)\\
	\Gamma_{\varphi\varphi\varphi\varphi}\,|_{\substack{ p=p_{\rm sym}(-\mu^2) \\ s=1}} 
            &=&-\lambda
\end{eqnarray}
lead to   
\begin{eqnarray}\label{bbRlam}
M\partial_M\frac{\partial^2}{\partial p^\rho\partial p^\sigma}
	    \Gamma_{\varphi\varphi h^{\mu\nu}}\,|_{\substack{ p=p_{\rm sym}(-\mu^2) \\ s=1}}
  +\beta^{\rm LZ}_{c_R}\frac{c_R}{2}(\delta^\mu_\rho\delta^\nu_\sigma+\delta^\mu_\sigma\delta^\nu_\rho)&=&0\\
	M\partial_M\Gamma_{\varphi\varphi\varphi\varphi}\,|_{\substack{ p=p_{\rm sym}(-\mu^2) \\ s=1}}
  +\beta^{\rm LZ}_\lambda(-\frac{\lambda}{4!})&=&0
\end{eqnarray}
Since at the test values none of the vertex functions depends on $M$, all
$\beta$-functions vanish too.\\

Hence at $s=1$ the LZ-equation 
\be\label{lzv}
M\partial_M \Gamma|_{s=1} =0
\ee
holds and reveals that the vertex functions, hence all other Green's
functions too, are independent of $M$ at $s=1$.

\subsection{The renormalization group equation}
The RG-equation formulates the response of the system to the variation of
the normalization parameter $\mu$, (see (\ref{highnorm})),
where e.g.\ couplings or field amplitudes are defined. Since the ST-operator does not depend on $\mu$ the partial differential operator
$\mu\partial_\mu$ is symmetric and can be expanded in the basis 
(\ref{tns}). Quite analogously to the LZ-equation (by removing Push and Shift)
we end up with 
\begin{eqnarray}\label{rg1}
	\mu\partial_\mu\Gamma_{|s=1}&=&
	(-\beta^{\rm RG}_{30} c_{30} \partial_{c_{30}}		      
	-\beta^{\rm RG}_{c_1} c_1 \partial_{c_1}  
	-\beta^{\rm RG}_{c_2} c_2 \partial_{c_2}
	-\beta^{\rm RG}_{c_R} c_R \partial_{c_R}	
    -\beta^{\rm RG}_\lambda \partial_\lambda\\
	&&+\gamma^{\rm RG}_\varphi\,\mathcal{N}_Y
	+\gamma^{\rm RG}_h\,\mathcal{N}_H
	+\gamma^{\rm RG}_c\,\mathcal{N}_L)\Gamma_{|s=1} \, .
\end{eqnarray}
We oberve that some normalization conditions involve $\mu$,
hence performing derivatives wrt $\mu$ does not commute with choosing
arguments for the relevant vertex functions and we expect non-trivial 
coefficient functions. 
Again we start with those tests which involve external fields.
The normalization condition  
\be\label{Yc1}
\frac{\partial}{\partial p^\lambda}\Gamma_{Y c^\rho \varphi}
      \,|_{\substack{ p=p_{\rm sym}(-\mu^2) \\ s=1} }
      =-i\eta_{\rho_\lambda}
 \ee
implies 
\be\label{Yc2}
\mu\partial_\mu\,\partial_p \Gamma_{Yc\varphi}\,|_{\substack{ p=p_{\rm sym}(-\mu^2) \\ s=1}}
-\gamma^{\rm RG}_\varphi(\partial_p \Gamma_{Yc\varphi}\,|_{\substack{ p=p_{\rm sym}(-\mu^2) \\ s=1}})=0 .
\ee
Now $\mu\partial_\mu$ does not commute with choosing a $\mu$-dependent 
argument, hence this determines $\gamma^{\rm RG}_\varphi$.

\be\label{Lcc2}
\frac{\partial}{\partial p^\lambda}\Gamma_{L_\rho c^\sigma c^\tau}
	       \,|_{\substack{ p=p_{\rm sym}(-\mu^2) \\ s=1}}	
      =-i\kappa(\delta^\rho_\sigma\eta_{\lambda\tau}
               -\delta^\rho_\tau\eta_{\lambda\sigma}).
\ee
Now $\mu\partial_\mu$ does not commute with choosing a $\mu$-dependent
argument, hence
\be
\mu\partial_\mu\frac{\partial}{\partial p^\lambda}
     \Gamma_{L_\rho c^\sigma c^\tau}
	       \,|_{\substack{ p=p_{\rm sym}(-\mu^2) \\ s=1}} 
      +i\gamma^{\rm RG}_c\kappa(\delta^\rho_\sigma\eta_{\lambda\tau}
               -\delta^\rho_\tau\eta_{\lambda\sigma})=0
\ee
which determines $\gamma^{\rm RG}_c$.
For the normalization condition
\be\label{Kc2}
\frac{\partial}{\partial p^\sigma}\Gamma_{K^{\mu\nu}c_\rho}
	                              \,|_{\substack{ p^2=-\mu^2 \\ s=1}}	
      =-i\kappa(\eta^{\mu\sigma}\delta_\rho^\nu+\eta^{\nu\sigma}\delta_\rho^\mu
                              -\eta^{\mu\nu}\delta_\rho^\sigma)
\ee
the structure is exactly the same as in the preceding
example such that the result is
\be
\mu\partial_\mu\frac{\partial}{\partial p^\sigma}\Gamma_{K^{\mu\nu}c_\rho}
	   \,|_{\substack{ p^2=-\mu^2 \\ s=1}}
+(\gamma^{\rm RG}_c-\gamma^{RG}_h)i\kappa(\eta^{\mu\sigma}\delta_\rho^\nu
+\eta^{\nu\sigma}\delta_\rho^\mu-\eta^{\mu\nu}\delta_\rho^\sigma)=0.
\ee
This equation gives $\gamma^{\rm RG}_h$.
The $\beta$-functions will be determined by the normalization conditions
for the couplings.
The normalization condition
\be\label{c3n}
\partial_{p^2}\gamma^{(2)}_{\rm TT}\,|_{\substack{ p^2=0 \\ s=1}} 
                             =c_{30}\kappa^{-2}
\ee
is independent from $\mu$ hence it implies
\be
\mu\partial_\mu \partial_{p^2}\gamma^{(2)}_{\rm TT}\,|_{\substack{ p^2=0 \\ s=1}}
              =0=-\beta^{\rm RG}_{30}c_{30}\kappa^{-2} 
      +2c_{30}\kappa^{-2}\gamma^{\rm RG}_h.
      \ee
This determines $\beta^{\rm RG}_{c_{30}}$.
The other normalization conditions, however depend on $\mu$ and thus result into   
\begin{eqnarray}\label{cmun}
\mu\partial_\mu 
\partial_{p^2}\partial_{p^2}\gamma^{(2)}_{\rm TT}\,|_{\substack{ p^2=-\mu^2 \\ s=1}}
	&=&2c_1\beta^{\rm RG}_1-2c_1\gamma^{\rm RG}_h\\
\mu\partial_\mu
\partial_{p^2}\partial_{p^2}\gamma^{(0)}_{\rm TT}\,|_{\substack{ p^2=-\mu^2 \\ s=1}}
	&=&-6c_2\beta^{\rm RG}_{c_2}+2c_1\beta^{\rm RG}_1c_1 
             +2(3c_2-c_1)\gamma^{\rm RG}_h\\
 \mu\partial_\mu\frac{\partial^2}{\partial p^\rho\partial p^\sigma}
	\Gamma_{\varphi\varphi h^{\mu\nu}}\,|_{\substack{p=p_{\rm sym}(-\mu^2) \\ s=1}}&=&
  (-\beta^{\rm RG}_{c_R}+2\gamma_\varphi+\gamma_h)c_R(\delta^\mu_\rho\delta^\nu_\sigma+\delta^\mu_\sigma\delta^\nu_\rho)\\
	\mu\partial_\mu\Gamma_{\varphi\varphi\varphi\varphi}\,|_{\substack{p=p_{\rm sym}(-\mu^2) \\s=1}}&=&
	(-\beta^{\rm RG}_\lambda+4\gamma^{\rm RG}_\varphi)\lambda
\end{eqnarray}
These equations determine 
$\beta^{\rm RG}_1,\beta^{\rm RG}_2,\beta^{\rm RG}_{c_{\rm R}},
\beta^{\rm RG}_\lambda$. They depend on the product $\mu\kappa$. Since we work 
in Landau gauge, they do not depend on a gauge parameter. \\

We now anticipate from ``finiteness properties'' that 
$\gamma^{\rm RG}_\varphi=\gamma^{\rm RG}_h=\gamma^{\rm RG}_c=0$ which implies that
also $\beta^{\rm RG}_{30}=0$. Hence if we would like to solve the RG-equation
for specific vertex functions we can start from
\be\label{cptRG}
(\mu\partial_\mu
+\beta^{\rm RG}_1c_1\frac{\partial}{\partial c_1}
+\beta^{\rm RG}_2c_2\frac{\partial}{\partial c_2}
+\beta^{\rm RG}_{c_{\rm R}}\frac{\partial}{\partial c_{\rm R}}
+\beta^{\rm RG}_\lambda\frac{\partial}{\partial \lambda})\gamma_{\rm fields}=0
\ee

The resulting linear partial differential equations are homogeneous and can 
be solved via characteristics. We define the variables
\begin{align}
t= ln(-\frac{p^2}{\mu^2}) \qquad u=\frac{c_3\kappa^{-2}}{p^2} 
\hspace{2mm} {\rm i.e.} \,\,\, ue^t=-\frac{c_3\kappa^{-2}}{\mu^2}
\end{align}
and the RG-equations become
\begin{align}
	(\frac{\partial}{\partial t}&
-\frac{1}{2}\beta_1(ue^t,g_1,g_2,g_{\rm R},g_\lambda)g_1\frac{\partial}{\partial g_1}
-\frac{1}{2}\beta_2(ue^t,g_1,g_2,g_{\rm R},g_\lambda)g_2\frac{\partial}{\partial g_2}\\
\nonumber
	&-\frac{1}{2}\beta_{\rm R}(ue^t,g_1,g_2,g_{\rm R},g_\lambda)g_{\rm R}\frac{\partial}{\partial g_{\rm R}}
	-\frac{1}{2}\beta_\lambda(ue^t,g_1,g_2,g_{\rm R},g_\lambda)g_\lambda\frac{\partial}{\partial g_\lambda})
	\gamma_{\rm fields}(e^t,e^tu,g_1,g_2,g_{\rm R},g_\lambda)=0. 
\end{align}
Here the running couplings $g_i \gets c_i, i=1,2,c_{\rm R},\lambda$ have 
to solve
\begin{align}\label{rcpls}
	\frac{dg_i}{dt}=-\frac{g_i}{2}\beta^{\rm RG}_i\,\, 
	i=1,2,c_{\rm R},\lambda\,\,({\rm no\, sum)}
\end{align}
(Prefactor and sign originate from using $\mu\partial_\mu$ instead of $\mu^2\partial_{\mu^2}$ and the explicit sign in the definition of $t$.)
Their solutions are given by
\begin{align}
g_i(t)=g_i(0)e^{-\frac{1}{2}\int^t_0 d\tau
	  \beta_i(ue^\tau,g_1(\tau),g_2(\tau),g_{\rm R}(\tau),g_\lambda(\tau))},
\end{align}
whereas
\begin{align}\label{tdpndu}
\frac{du}{dt}=0 \Rightarrow u(t,u_0,g_i(0))=u_0.
\end{align}

The most important examples in the following will be the
$r=2,0; L=K=T$ components of $\Gamma_{hh}$ (The solution for other vertex functions
runs along similar lines.) According to naive dimensional
considerations one first makes explicit factors $p^2p^2$, then one separates
tree contributions from higher orders. The result is given by
\begin{align}\label{hhtgmm}
	\gamma^{(2)}_{\rm TT}=&p^2p^2\hat{\hat{\gamma}}^{(2)}=
	p^2c_3\kappa^{-2}+p^2p^2\hat{\hat{\gamma}}^{(2)}(e^t,u_0,g_i(t))\,\,
	{\rm all}\,\, i\\
	\gamma^{(0)}_{\rm TT}=&p^2p^2\hat{\hat{\gamma}}^{(0)}=
-2p^2c_3\kappa^{-2}+p^2p^2\hat{\hat{\gamma}}^{(0)}(e^t,u_0,g_i(t))\,\,
       {\rm all}\,\, i
\end{align}
This separation  determines the starting points of 
$\hat{\hat{\gamma}}$:
\begin{eqnarray}
\hat{\hat{\gamma}}^{(2)}_{\rm TT}(t,u_0,g_i)|_{t=0}&=&-g_1(0)
	     =-c_1\\
\hat{\hat{\gamma}}^{(0)}_{\rm TT}(t,u_0,g_i)|_{t=0}&=&3g_2(0)+g_1(0)
            =3c_2+c_1
\end{eqnarray}
Obviously the solutions become trivial when putting $c_1=c_2=0$.\\
A further restriction originates from the scheme, which fixes 
$\partial_{p^2}\gamma^{(2)}_{\rm TT}$ at $p^2=0,s=1$ and accordingly we are
not allowed to admit counterterms which would change this.
Hence the dependence of $\hat{\hat{\gamma}}^{(r)}$ from $u=-c_3\kappa^{-2}/p^2$ is restricted to the value $u_0=-c_3\kappa^{-2}/\mu^2$.
Actually, this is just the content of (\ref{tdpndu}).
It fits to the fact, that $c_3\kappa^{-2}$ does not run.\\ 
The interpretation is as follows: the $c_3\kappa^{-2}$ terms are tree values which are not corrected by higher orders.
The $\hat{\hat{\gamma}}$ terms provide for the value $t=0$ the tree approximation contributions going with $c_1,c_2$. For $t> 0$ they comprise all higher order corrections expressed in terms of the running couplings.
One should note that these results hold at $s=1$, the physical value.\\
We underline, by repeating: the separation in tree, resp.\ higher order 
contributions in $\gamma^{(r)}_{\rm TT}, r=0,2$, $(\ref{hhtgmm})$, together with 
the non-renormalization of $\kappa^{-2}$ and $c_3$ just means that only the higher 
derivatives and the scalar interactions are responsible for the running of 
couplings, i.e.\ of $c_1,c_2,c_{\rm R},\lambda$.
In this respect the EH part is only a kind of spectator.\\
It is appropriate at this point to have a look back to \cite{EKKSI,EKKSII}. There it was
possible and interesting to reduce the coupling $\lambda$ to $c_{\rm R}$, i.e.\
to express $\lambda$ as a function of $c_{\rm R}$ consistent with renormalization.
As a matter of fact this is no longer possible in the wider framework discussed 
here. The gravity selfinteraction prevents such a dependence.\\ 

The common understanding of running couplings and their use in phenomenology
(QCD, electroweak standard model) is that inserting
them in place of a tree coupling at a given order in perturbation theory
``improves'' the results of that order, i.e.\ in some qualitative sense 
extends those to all orders.\\
For the model under consideration,
in the literature mostly an expansion in terms of $\kappa^2$ has been performed.
This we do not do because the renormalization of the electroweak standard model
teaches us an important lesson.
If one wants to ensure there poles for physical particles one has to use on-shell 
normalization conditions.
But then the couplings have to be realized as mass ratios.
This in turn implies that even their $\beta^{\rm CS}$-functions can no longer be
expanded as power series in the couplings, but only in the number of loops 
\cite{kraus1999callan}. (CS-$\beta$ functions are, as a rule, simpler than those
of the RG equation.)
The reason for this is the same as here: they are complicated polylogarithmic functions of the couplings.
Hence an expansion in terms of $\kappa^2$, here, may very well be misleading.
E.g.\ the pure fact that after using such an expansion the $\beta$-functions
come out as rational functions in the couplings is suspicious. Relying on
this outcome and concluding from there on the asymptotic behaviour seems
to be courageous.\\

\section{Finiteness Properties}
%file finitelandau.tex\\
It is well-known [BPS91] that in ordinary pure Yang-Mills theory the anomalous dimension of the vector field vanishes as well as that of the Faddeev-Popov ghost $c$ when working in Landau gauge.
In papII we extended this result by showing that the analogue is true for the 
fields $h^{\mu\nu}, c^\mu$. There it implied that $\beta_3$ also vanishes.
In the present paper we shall see that correspondingly $\gamma_\varphi=0$.\\
Like in papII one starts from the integrated antighost equation of motion
\begin{align}\label{aghst}%anti-ghost eq.
	\int\frac{\delta\Gamma}{\delta c^\mu}\equiv& 
\int(\frac{\delta\Gamma_{\phi\pi}}{\delta c^\mu}
	+\frac{\delta\Gamma_{\rm ext.f.}}{\delta c^\mu})\\ 
	=&\int(-\kappa(\frac{1}{2}D^{\mu\nu}
        (\partial_\mu\bar{c}_\nu+\partial_\nu\bar{c}_\mu)
	-K_{\mu\nu} \partial_\rho h^{\mu\nu}+L_\lambda\partial_\rho c^\lambda)
	-\frac{3}{4}Y\partial_\mu\varphi+\frac{1}{4}\partial_\mu Y\varphi)
\end{align}
and combines it with the gauge condition to form
\begin{align}\label{fntnc}%finiteness condition
\bar{\mathcal{G}}\Gamma\equiv \int(\frac{\delta\Gamma}{\delta c^\rho}
+\kappa\partial_\rho\bar{c}_\lambda\frac{\delta\Gamma}{\delta b_\lambda})
	= \int(\kappa(K_{\mu\nu}\partial_\rho h^{\mu\nu} 
	-L_\lambda\partial_\rho c^\lambda)
	-\frac{3}{4}Y\partial_\rho\varphi+\frac{1}{4}\partial_\rho Y\varphi).
\end{align}
Since this expression is linear in the quantized fields it can be naively extended
to all orders in the form as it arises in tree approximation (s.\ papII Appendix,
$A_1$) for more details).\\
Potential counterterms, one could have been obliged to add, must be independent 
of $b_\mu$, could depend on $\bar{c}$ only via $H_{\mu\nu}$ and must satisfy 
the ghost equation
\begin{align}\label{ghqt}%ghost equation
\kappa\frac{\delta\bar{\Gamma}}{\delta\bar{c}_\mu}
+\partial_\lambda\frac{\delta\bar{\Gamma}}{\delta K_{\mu\lambda}}=0.
\end{align}
The candidates for this are given by 
\begin{align}\label{ssrtL} %symmetric insertion L
\Delta_L\cdot\Gamma=f_L(\alpha_0)\mathcal{N}_L\Gamma-\chi 
       f\prime_L\int Lc=-\mathcal{B}_{\bar{\Gamma}}(f_L\int Lc)\qquad
\mathcal{N}_L\equiv \int c\frac{\delta}{\delta c}
	            -\int L\frac{\delta}{\delta L}
\end{align}
In Landau gauge $\alpha_0=0, \chi=0$, hence $f_L$ is a number.
To be satisfied is (\ref{fntnc}), but 
\begin{align}\label{chck1}%check1
\Delta_L\cdot\Gamma=-\mathcal{B}_{\bar{\Gamma}}(\int Lc)=
  \kappa\int L_\rho c^\lambda\partial_\lambda c^\rho + L-{\rm independent},
 \end{align}
would contribute to the rhs of (\ref{fntnc}) a term 
$\int L_\lambda\partial_\rho c^\lambda$,
which would, however, change the coefficient of the term already present. This
is forbidden, hence this term is excluded as a counterterm, the field $c$ is 
not renormalized: $\gamma^{\rm RG}_c=0$. The next
candidate is $\Delta_H$.
\begin{align}\label{ssrtH} %symmetric insertion H
\Delta_H\cdot\Gamma=f_H\mathcal{N}_K\Gamma
 +\chi f^\prime_H(\alpha_0)(\int(Kh
      -\bar{c}\frac{\delta\Gamma_{\rm class}}{\delta b}) 
              +2\alpha_0\frac{\partial}{\partial\chi}\Gamma_{\rm class}
	 \qquad     \mathcal{N}_K\equiv N_h-N_K-N_b
\end{align}
Again, in Landau gauge $\alpha_0=0, \chi=0$, $f_H$ is a number.
\be\label{ssrT2}
\Delta_H = f_H\int(-\kappa K_{\mu\nu}D^{\mu\nu}_\rho c^\rho)
\ee
This would contribute to the rhs of (\ref{fntnc}) a term
$\int \kappa K_{\mu\nu}\partial_\rho h^{\mu\nu}$, again a term which is 
already present and
whose coefficient must not be changed. So, this counterterm, too, is forbidden.
The field $h^{\mu\nu}$ is not renormalized: $\gamma^{\rm RG}_h=0$,
hence, in view of the normalization conditions, $\beta^{\rm RG}_3=0$.\\
As further, matter dependent contribution, we have identified 
\be\label{ssm}%symmetric insertion matter
\Delta_Y\cdot\Gamma=\mathcal{B}_{\bar{\Gamma}}(f_Y\int Y\varphi)
=f_Y\mathcal{N}_Y\Gamma+\chi f\prime(\alpha_0)(\int Y\varphi)
 \qquad \mathcal{N}_Y\equiv N_\varphi-N_Y
\ee
In Landau gauge $f_Y$ is a number.
\be\label{ssm2}
\Delta_Y\cdot \Gamma= f_Y\int Y(c^\lambda\partial_\lambda\varphi
   +\frac{1}{4}\partial_\lambda c^\lambda \varphi)
\ee
As a counterterm this would add to (\ref{fntnc}) a term 
$-\frac{3}{4}Y\partial_\mu+\frac{1}{4}\partial_\mu Y\varphi$ which also is
already present, hence forbidden.\\
This implies $\gamma^{\rm RG}_\varphi=0$, the field amplitude of $\varphi$ is not
renormalized in Landau gauge.\\

\section{Rigid Weyl Invariance}
%file rigWeyl.tex\\
From  \cite{PS23} we know that a proper substitute of scaling in flat
spacetime, described there by the Callan-Symanzik equation, is rigid Weyl
invariance in curved spacetime. Its ST-symmetric functional differential
operator on $\Gamma$ reads 
\be\label{rgWprt}%rigid Weyl operator
\frac{1}{2\sigma}W^{\rm W}_{\rm rig}\equiv \int(
  g^{\mu\nu}\frac{\delta}{\delta g^{\mu\nu}}
 -K_{\mu\nu}\frac{\delta}{\delta K_{\mu\nu}}
 -\bar{c}_\mu\frac{\delta}{\delta \bar{c}_\mu}
 -b_\mu\frac{\delta}{\delta b_\mu}),
 \ee
because the field $\varphi$ and hence also its counting operator 
$\mathcal{N}_Y$ are invariant under rigid Weyl transformations, hence do not 
show up here.\\
It leads to
\begin{align}\label{rgWqt}%rigid Weyl equation
W^{\rm W}_{\rm rig}\Gamma_{|s=1}=&-2\sigma\hat{\alpha}c_3\kappa^{-2}\lbrack
\int\,\sqrt{-g}R\rbrack^4_4\cdot\Gamma_{|s=1}\\
                           =&-2\sigma\lbrace\hat{\alpha}c_3\kappa^{-2}
	    \lbrack\int\,\sqrt{-g}R\rbrack^3_3\cdot\Gamma_{|s=1}\\
	&+\lbrack\int\,(\sqrt{-g}(u_1R^{\mu\nu}R_{\mu\nu}+u_2R^2)
	+u_R(-g)^{1/4}\varphi^2R+u_\lambda \varphi^4)\rbrack^4_4\\
	&+u_h\mathcal{N}_H+u_c\mathcal{N}_L+u_\varphi\mathcal{N}_Y\rbrace
	   \cdot\Gamma_{|s=1}
\end{align}
Here we have used appropriate Zimmermann identities in order to replace
the hard EH-insertion by its soft partner and the respective symmetric hard 
corrections.\\
The $\lbrack\cdot\cdot\cdot\rbrack^4_4$ insertions can be replaced 
by the symmetric derivatives with respect to their couplings
\begin{align}\label{rgWcpls}%rigid Weyl couplings
\frac{1}{-2\sigma}W^{\rm W}_{\rm rig}\Gamma_{|s=1}=&\hat{\alpha}c_3\kappa^{-2}
	    \lbrack\int\,\sqrt{-g}R\rbrack^3_3\cdot\Gamma_{|s=1}\\
	&+\lbrace u_1\frac{\partial}{\partial c_1}
	         +u_2\frac{\partial}{\partial c_2}
		 +u_R\frac{\partial}{\partial c_R}
           +u_\lambda\frac{\partial}{\partial \lambda}\\
	&+u_h\mathcal{N}_H+u_c\mathcal{N}_L+u_\varphi\mathcal{N}_Y\rbrace
	   \Gamma_{|s=1}
\end{align}
This form clearly shows that the rigid Weyl identity plays the role of
a Callan-Symanzik equation: the soft $\lbrack...\rbrack^3_3\cdot\Gamma$
vanishes in the deep Euclidian region, the $u\partial$ terms 
correspond to $\beta\partial$ and the leg counting operators
$\mathcal{N}$ appear with the anomalous dimensions $u$ as factors.
In the present context $u_h,u_c,u_\varphi$ correspond to anomalous
Weyl weights of the fields $h,c,\varphi$.\\

It is also interesting to explore what this equation implies on-shell,
i.e.\ after projection to Fock space. One first goes over via Legendre
transformation to connected Green's functions $Z_c(\underbar{J})$, then
by exponentiation to general Green's functions $Z=\exp(iZ_c)$.
\begin{align}\label{wrigWG}%W-id rig Weyl for General
\frac{1}{-2\sigma}W^{\rm W}_{\rm rig}(\underbar{J})Z_{|s=1} 
=&\hat{\alpha}c_3\kappa^{-2}
	    \lbrack\int\,\sqrt{-g}R\rbrack^3_3\cdot Z_{|s=1}\\
	&+\lbrace u_1\frac{\partial}{\partial c_1}
	         +u_2\frac{\partial}{\partial c_2}
		 +u_R\frac{\partial}{\partial c_R}
	   +u_\lambda\frac{\partial}{\partial \lambda}
	&+u_h\mathcal{N}_H+u_c\mathcal{N}_L+u_\varphi\mathcal{N}_Y\rbrace
	   Z_{|s=1}
\end{align}
The application of the projector $:\Sigma:$ from (\ref{prjfr})  changes
the lhs into rigid Weyl transformations of the quantum in-fields applied to
the $S$-operator, whereas the soft term on the rhs becomes its operator
equivalent, the derivatives wrt the couplings act on the $S$-operator
and the number operators $\mathcal{N}$ are projected to zero.
\begin{align}\label{wrigSop}%W-id rig Weyl for S-operator
	\frac{1}{-2\sigma}W^{\rm W}_{\rm rig} 
 \equiv&\int\lbrace z^{-1}_\phi \phi^{\mu\nu}\frac{\delta}{\delta \phi^{\mu\nu}}
		   +z^{-1}_{\bar{c}}\bar{c} \frac{\delta}{\delta\bar{c}}
		   -z^{-1}_b b\frac{\delta}{\delta b}\rbrace \\
   \frac{1}{-2\sigma}W^{\rm W}_{\rm rig}S^{\rm op}= 
	&\hat{\alpha}c_3\kappa^{-2}(\lbrack\int\,\sqrt{-g}R\rbrack^3_3)^{\rm op}\\
	&+(u_1\frac{\partial}{\partial c_1}
	         +u_2\frac{\partial}{\partial c_2}
		 +u_R\frac{\partial}{\partial c_R}
		 +u_\lambda\frac{\partial}{\partial\lambda})S^{\rm op}
\end{align}
Several comments are in order. As compared with (\ref{rgWprt}) we went over
from $g^{\mu\nu}=\eta^{\mu\nu}+h^{\mu\nu}$ to $h^{\mu\nu}$. Then, recalling
that in the two-field-approximation of $h^{\mu\nu}$ those are to be described
by the fields $\phi^{\mu\nu},\Sigma^{\mu\nu}$, we used that 
$\delta^{\rm W}_{\rm rig}\phi= 2\sigma \phi^{\mu\nu}$,
however $\delta^{\rm W}_{\rm rig}\Sigma^{\mu\nu}=0$. Here all fields are 
free quantum ``in''
fields as they show up in the projector $:\Sigma:$. The factors $z^{-1}$
are the residues of the respective propagators.\\
There is no explicit appearance of the matter field $\varphi$ because it is
invariant under rigid Weyl.\\

\section{No massive higher order zero's}\label{sec:abshighzero}
%file nohigherorderz.tex\\

Most important for the physical interpretation of the model is the understanding 
of the zero's of $\gamma^{(r)}_{\rm TT}$.
The one's at $p^2=0$ are fixed, guaranteed by the scheme and RG invariant: 
they are physical.
But the second zero's
can not be continued to higher orders as we shall show now.\\
We consider the case $r=2$ up to and including one-loop.
\be\label{sp2gmm}%spin2gamma
(\gamma^{(2)}_{\rm TT})^{(\le 1)}_{|s=1}=p^2c_3\kappa^{-2}-c_1p^2p^2
-c^{(1)}_1p^2p^2+p^2p^2(\hat{\hat{\gamma}}^{(2)}_{\rm TT})_{|\rm nt}
\ee
Here $c^{(1)}_1$ is the coefficient of the one-loop counterterm
to the invariant $\int\sqrt{-g}R^{\mu\nu}R_{\mu\nu}$; nt means
``non-trivial''
i.e.\ these are the contributions of the non-trivial diagrams in one-loop
order (the counterterm is pointlike, hence a trivial diagram). The first
zero at $p^2=0$ is obvious. We claim that
\be\label{2zgmm}%second zero
0=c_3\kappa^{-2}-c_1p^2-c^{(1)}_1p^2
            +p^2(\hat{\hat{\gamma}}^{(2)}_{\rm TT})_{|\rm nt}
\ee
has no solution for $p^2=c_3\kappa^{-2}/c_1$  and the counterterm 
coefficient with its value as given by the normalization condition for $c_1$
\begin{align}\label{cttc}%counterterm coefficient
c^{(1)}_1&=\frac{1}{2}\frac{\partial}{\partial p^2}\frac{\partial}{\partial p^2}
(\gamma^{(2)}_{\rm TT})^{(1)}_{|p^2=-\mu^2,s=1}\\
	&=(\hat{\hat{\gamma}}^{(2)}_{\rm TT})^{(1)}_{|p^2=-\mu^2,s=1}
	+\lbrack (2p^2\partial_{p^2}
	+\frac{1}{2}p^2p^2\partial_{p^2}\partial{_p^2})
	                  (\hat{\hat{\gamma}}^{(2)}_{\rm TT})^{(1)}
	\rbrack_{|p^2=-\mu^2,s=1}.
\end{align}
Hence (\ref{2zgmm}) boils down to 
\be\label{xct}%existence condition
\frac{c_3\kappa^{-2}}{c_1}(-c^{(1)}_1
 +(\hat{\hat{\gamma}}^{(2)}_{\rm TT})_{|\rm nt})=0,
\ee
with the arguments of $\gamma$ being 
$(-p^2/\mu^2,c_3\kappa^{-2}/p^2,c_1,c_2,c_{\rm R},\lambda) \rightarrow  
(-c_3\kappa^{-2}/(c_1\mu^2),c_1,c_1,c_2,c_{\rm R},\lambda) $. 
More explicitly
\begin{align}\label{ctdt}%contradiction
-\left(\hat{\hat{\gamma}}^{(2)}_{\rm TT})^{(1)}
	(1,-\frac{c_3\kappa^{-2}}{c_1\mu^2},c_1,c_2,c_{\rm R},\lambda)
	+\lbrack (-2\mu^2\partial_{p^2}
	+\frac{1}{2}\mu^2\mu^2\partial_{p^2}\partial{_p^2})
	              (\hat{\hat{\gamma}}^{(2)}_{\rm TT})^{(1)}
	\rbrack_{|p^2=-\mu^2,s=1}\right)_{|\rm nt}&\\
+\left(\hat{\hat{\gamma}}^{(2)}_{\rm TT}
(-\frac{c_3\kappa^{-2}}{c_1\mu^2},c_1,c_1,c_2,c_{\rm R},\lambda)\right)_{|\rm nt}
	 =&0,
\end{align}
all taken at $s=1$. (In this explicit form also the first bracket refers to
the non-trivial diagrams.) It is to be noted that for the $\gamma$ in the 
first line
the  $p^2$-argument is at an unphysical value, whereas for the $\gamma$ in
the second line it is at a physical point. Therefor
this equation can not be satisfied. Hence beginning with one loop the
respective propagator, $<hh>^{(2)}_{\rm TT}$,  has no second pole. Obviously 
this is also true for the case $r=0$.\\
Hence, this argument from papII holds completely unchanged also in the present 
case, where in addition to all other fields of EH + hds
we have an interacting matter field: the $\gamma$'s in question
just depend also on the couplings $c_{\rm R}$ and $\lambda$, but this dependence 
does not change the outcome.\\

\section{Projection to physical state space}
\subsection{The single pole fields} 
%file: lagrangemult\\

The field $h^{\mu\nu}$ which we used up to now has double poles, hence
is not suited for the construction of a conventional Fock space. The idea,
due to Stelle, is  to decompose its bilinear contributions in field space
into those of single pole fields, $\phi^{\mu\nu},\Sigma^{\mu\nu}$, for which  
ordinary Fock spaces can be constructed and to describe eventually scattering 
etc in the tensor product of these two spaces. Interaction should still be
described in terms of the double pole field $h$, such that all results obtained
up to now can be taken over, however truely physical, on-shell quantities
always require additional treatment.\\
Starting point is the decomposition of the $h^{\mu\nu}$ propagators into partial 
fractions which have only simple poles, as presented in 
\cite[eqs. $C_1,C_2$]{pottel2021perturbative}.
\begin{align}\label{dcprgt}%decomposition of propagators
\langle hh \rangle^{(2)}_{\rm TT}&=
\frac{-i}{p^2-m^2}\cdot\frac{1}{c_1p^2-c_3\kappa^{-2}}	=
	\frac{1}{c_3\kappa^{-2}-c_1m^2}\cdot\frac{i}{p^2-m^2}
	+\frac{-i}{c_3\kappa^{-2}(p^2-\frac{c_3\kappa^{-2}}{c_1})}\\ 
\langle hh \rangle^{(0)}_{\rm TT}&=
	\frac{i}{p^2-m^2}\cdot\frac{1}{(3c_2+c_1)p^2-2c_3\kappa^{-2}}=
	\frac{1}{2c_3\kappa^{2}+2(3c_2+c_1)m^2}\cdot\frac{i}{p^2-m^2}\\ \nonumber
&-\frac{1}{2c_3\kappa^{-2}}\cdot\frac{i}{p^2+\frac{c_3\kappa^{-2}}{2(3c_2+c_1)}}
\end{align}
We are looking for fields $\phi^{\mu\nu},\Sigma^{\mu\nu}$ whose bilinear terms in the action just yield these simple pole propagators: $\phi$ the massless, $\Sigma$ the massive ones.
With this aim in mind one decomposes the field-bilinear part of the classical invariants of EH + hds with the help of a Lagrange multiplier $Z^{\mu\nu}$ such that only second derivatives of $h^{\mu\nu}$ and $Z^{\mu\nu},\Sigma^{\mu\nu}$ respectively show up.
Since in \cite{stelle1978classical} in an analogous context this problem has been solved we can proceed the other way round: we start from
\begin{equation}\label{hdcpt}
h^{\mu\nu}=\phi^{\mu\nu}+\Sigma^{\mu\nu}\qquad
Z^{\mu\nu}=\phi^{\mu\nu}-\Sigma^{\mu\nu}
\end{equation}
as desired field decomposition and from
\begin{align}\label{dcpta}
	\Gamma(\phi)&=\Gamma_{\rm EH}(\phi)\\
	\Gamma(\Sigma)&=-\Gamma_{\rm EH}(\Sigma)+\Gamma_{\rm mass}(\Sigma)\\
	\Gamma_{\rm EH}(\phi)&=\frac{\tilde{c}_3\kappa^{-2}}{4}\int(
-\phi^{\mu\nu}\Box\phi_{\mu\nu}+\phi^\rho_\rho\Box\phi^\sigma_\sigma
-2\phi^{\mu\nu}\partial_\mu\partial_\nu\phi^\lambda_{\phantom{\lambda}\lambda}
	+2\phi^{\mu\nu}\partial_\rho\partial_\nu\phi^\rho_\mu)    \\
	-\Gamma_{\rm EH}(\Sigma)+\Gamma_{\rm mass}&=\frac{\tilde{c}_3\kappa^{-2}}{4}\int(
\Sigma^{\mu\nu}\Box\Sigma_{\mu\nu}-\Sigma^\rho_\rho\Box\Sigma^\sigma_\sigma
+2\Sigma^{\mu\nu}\partial_\mu\partial_\nu\Sigma^\lambda_{\phantom{\lambda}\lambda}
	-2\Sigma^{\mu\nu}\partial_\rho\partial_\nu\Sigma^\rho_\mu    \\
	&\phantom{\frac{c_3\kappa^{-2}}{4}\int)}
        +a_2\Sigma^{\mu\nu}\Sigma_{\mu\nu}
	+a_0(\Sigma^\lambda_{\phantom{\lambda}\lambda})^2)   \nonumber 
\end{align}
as desired bilinear action in order to identify at a convenient stage in our conventions the mass $a_2,a_0$ and coupling $\tilde{c}_3\kappa^{-2}$ parameters. 
The relative minus sign of the two actions just represents the negative residue sign of massive propagators in \cite[eqs. $C_1,C_2$]{pottel2021perturbative}.
As an aside we note that the mass term is not of Fierz-Pauli type, since it will turn out that $a_2+a_0\not=0$.
Hence it contains some spin $0$ component.
(The Fierz-Pauli condition $a_2+a_0=0$ would remove within $\gamma^{(0)}_{\rm TT}$ the second zero, hence ruin UV convergence.)\\
For the subsequent treatment we give here the actions in projector form 
(s.\ Appendix).
\begin{align}\label{prjctfrm}
	\Gamma_{\rm EH}(\phi)=\frac{\tilde{c}_3\kappa^{-2}}{4}\int(
&-\phi^{\mu\nu}\Box(P^{(2)}_{\rm TT}+P^{(1)}_{\rm SS}
                  +P^{(0)}_{\rm TT}+P^{(0)}_{\rm WW})\phi^{\rho\sigma}\\
&+\phi^{\mu\nu}\Box(3P^{(0)}_{\rm TT}+\sqrt{3}(P^{(0)}_{\rm TW}
		  +P^{(0)}_{\rm WT})+P^{(0)}_{\rm WW})\phi^{\rho\sigma}\\
&-\phi^{\mu\nu}\Box(\sqrt{3}(P^{(0)}_{\rm TW}+P^{(0)}_{\rm WT})
	            +2P^{(0)}_{\rm WW})\phi^{\rho\sigma}\\
	&+\phi^{\mu\nu}\Box(P^{(1)}_{\rm SS}
		    +2P^{(0)}_{\rm WW})\phi^{\rho\sigma})
\end{align}
\begin{align}\label{prjct2}
	\Gamma(\Sigma)=&-\Gamma_{\rm EH}(\Sigma)+\Gamma_{\rm mass}(\Sigma)\\
	=&\frac{\tilde{c}_3\kappa^{-2}}{4}\int(
	\Sigma^{\mu\nu}(\Box+a_2)(P^{(2)}_{\rm TT}+P^{(1)}_{\rm SS}
                  +P^{(0)}_{\rm TT}+P^{(0)}_{\rm WW})\Sigma^{\rho\sigma}\\
 &-\Sigma^{\mu\nu}(-\Box+a_0)(3P^{(0)}_{\rm TT}+\sqrt{3}(P^{(0)}_{\rm TW}
		  +P^{(0)}_{\rm WT})+P^{(0)}_{\rm WW})\Sigma^{\rho\sigma}\\
&+\Sigma^{\mu\nu}\Box(\sqrt{3}(P^{(0)}_{\rm TW}+P^{(0)}_{\rm WT})
	            +2P^{(0)}_{\rm WW})\Sigma^{\rho\sigma}\\
&-\Sigma^{\mu\nu}\Box(P^{(1)}_{\rm SS}
		    +2P^{(0)}_{\rm WW})\Sigma^{\rho\sigma})
\end{align}
In the next step we replace the fields: $\phi=h+Z, \Sigma=h-Z$ and go over to a total action
\begin{align}\label{tctn}%total action
\Gamma(\phi)+\Gamma(\Sigma)=\Gamma_{\rm total}\rightarrow \Gamma(h,Z)
\end{align}
We find %10.5.2022 p.9
\begin{align}\label{hZfctn}%h,Z form of action
	\Gamma_{\rm total}(h,Z)=\frac{\tilde{c}_3\kappa^{-2}}{4}\int(
	&-2h(P^{(2)}_{\rm TT}(2\Box+a_2)+P^{(1)}_{\rm SS}a_2
	 +P^{(0)}_{\rm TT}(-4\Box+a_2+3a_0)\\
	& +\sqrt{3}(P^{(0)}_{\rm TW}
	 +P^{(0)}_{\rm WT})a_0+P^{(0)}_{\rm WW}(a_2+a_0))Z\\
        &+h(P^{(2)}_{\rm TT}a_2+P^{(1)}_{\rm SS}a_2+P^{(0)}_{\rm TT}(a_2+3a_0)\\
	&+(\sqrt{3}P^{(0)}_{\rm TW}+P^{(0)}_{\rm WT})a_0+P^{(0)}_{\rm WW}(a_2+a_0)
	 )h\\
        &+Z(P^{(2)}_{\rm TT}a_2+P^{(1)}_{\rm SS}a_2+P^{(0)}_{\rm TT}(a_2+3a_0)\\
	&+(\sqrt{3}P^{(0)}_{\rm TW}+P^{(0)}_{\rm WT})a_0+P^{(0)}_{\rm WW}(a_2+a_0)
	 )Z
 \end{align}
This action has the desired structure 
$\int(h\mathcal{D}_{hZ}Z+hM_{\rm hh}h+ZM_{ZZ}Z)$ with $Z$ representing the 
Lagrange multiplier field.
This explicit form has not been presented in \cite{stelle1978classical}.\\
The final form $\Gamma(h)$, which can be compared with EH+hds, is now obtained by eliminating $Z$ via its equation of motion
\begin{align}\label{qtnmZ}%equation of motion for Z
\frac{\delta \Gamma}{\delta Z}(h,Z)=0.
\end{align}
One obtains
\begin{align}\label{rstZH}%result ZH
	(P^{(2)}_{\rm TT}a_2	
	+P^{(1)}_{\rm SS}a_2	
	+P^{(0)}_{\rm TT}(a_2+3a_0)	
	+\sqrt{3}(P^{(0)}_{\rm TW}+P^{(0)}_{\rm WT})a_0	
	+P^{(0)}_{\rm WW}(a_2+a_0))Z\\	
	=(P^{(2)}_{\rm TT}(2\Box +a_2) 	 
	+P^{(1)}_{\rm SS}a_2	
	+P^{(0)}_{\rm TT}(-4\Box+a_2+3a_0)\\	
	+\sqrt{3}(P^{(0)}_{\rm TW}+P^{(0)}_{\rm WT}a_0	
	+P^{(0)}_{\rm WW}(a_2+a_0))h	
\end{align}
Suitably projecting and  equating coefficients one can solve for $Z$ in terms of $h$.
Inserting into (\ref{hZfctn}) one arrives finally at
\begin{align}\label{frth}%final result in terms of h
	\Gamma_{\rm total}(h)=\frac{\tilde{c_3}\kappa^{-2}}{4}\int h
   (P^{(2)}_{\rm TT}(-\frac{4}{a_2}\Box(\Box+a_2))
	+P^{(0)}_{\rm TT}(-\frac{8}{a_2+3a_0}
	                \Box(\Box+(a_2+3a_0))))h 
\end{align}
This result permits identification of the parameters:
\begin{align}\label{prntfn}
	\frac{1}{4}\tilde{c}_3\kappa^{-2}=c_3\kappa^{-2} 
 \qquad             a_2=\frac{4c_3\kappa^{-2}}{c_1}
\qquad a_0=-\frac{3c_2+2c_1}{3c_1(3c_2+c_1)}c_3\kappa^{-2} 
\end{align}
\\
%\P\P This identification has to be verified! A problematic factor 4 can easily be restored: one just has to note that the inverse relation to (ref{hddcpt}) has a factor $\frac{1}{2}$ for $\phi,\Sigma$ in terms of $h,Z$. S. early calculations 7.6.2022, 10.5.2022 p,7,12
%\P\P 

\subsection{Projection to Einstein-Hilbert} \label{sec:projectionEH}
%file: projection.tex\\
We start from the model constructed to all orders in
\cite{pottel2021perturbative} in terms of the double pole field $h^{\mu\nu}$ 
and indicate now, how to identify the fields $\phi^{\mu\nu},\Sigma^{\mu\nu}$ and 
their use within that given model.\\
The massive field $\Sigma^{\mu\nu}$ will beginning with one loop no longer 
refer to the propagation of a particle: its propagator could have at the very best
a complex pole. (Due to the properties of the polylogs it could also have 
another singular character.) However as we have seen above this would-be
pole can not be reached -- beginning with one-loop. There are no parameters 
available which could in accord with the $s$-symmetry protect the real part of the 
possible singularity from being shifted in higher orders. This is the
meaning of the non-invariance under RG in accordance 
with $(\ref{2zgmm})$. This is a clear hint that it is unphysical, apart from its
negative norm properties in the tree approximation. We continue this discussion
after having described the projection procedure to the physical Hilbert space.\\

We identify the massless spin two field
$\phi^{\mu\nu}$ with the massless spin two graviton field, together with the 
fields $c,\bar{c},b$ as companions for building up the Kugo-Ojima \cite{KuOj} 
doublets. We can proceed for $\phi$ this way because it satisfies all 
requirements which one expects for such a field. Is has the correct 
covariance under $s$ and can in all respects be derived from
\cite{pottel2021perturbative}: one replaces there $h^{\mu\nu}$ within 
$\Gamma_{\rm eff}=\Gamma^{\rm class}+\Gamma^{\rm countert}$
by $h^{\mu\nu}=\phi^{\mu\nu}+\Sigma^{\mu\nu}$ in the field expansion 
with number of fields $n$ greater or equal to three.
(In particular for the counterterms too $h=\phi+\Sigma\,$.)
In tree approximation bilinear terms and in gauge fixing, Faddeev-Popov and 
external field terms, $h^{\mu\nu}$ is simply replaced by $\phi^{\mu\nu}$, whereas 
the field $\Sigma^{\mu\nu}$ comes along with the terms given in (\ref{dcpta}) 
(upon replacing the mass parameters $a_2,a_0$ with their values given in 
(\ref{prntfn})).\\
The general Green's functions in terms of $h$ give rise to those of 
$\phi^{\mu\nu},\Sigma^{\mu\nu}$ for number of fields $n$ greater or equal to 
three by 
introducing respective sources $j_\phi,j_\Sigma$, fitting to $h=\phi+\Sigma$.
We now go over to the $S$-operator by projecting
general Green's functions $Z(\underbar{J})$ down to Fock space.
The fields appearing in the projector are free in-fields and related to their 
corresponding wave function operators $K$
\be\label{prjfr}
	S^{\rm op}=:\Sigma:Z(\underbar{J})_{|{\underbar{J}=0}}\qquad
	:\Sigma:= :\exp{(X)}:
\ee
\begin{align}
X\equiv \int dxdy (
	&\phi^{\mu\nu}(x)K^{\phi\phi}_{\mu\nu\rho\sigma}(x-y)z^{-1}
	\frac{\delta}{\delta j^{\phi}_{\rho\sigma}}(y)
	+\phi^{\mu\nu}(x)K^{\phi b}_{\mu\nu\rho}(x-y)z^{-1}
	\frac{\delta}{\delta j^b_\rho}(y)\\
	+&b^\rho(x)K^{b\phi}_{\rho\alpha\beta}(x-y)z^{-1}	
	\frac{\delta}{\delta j^\phi_{\alpha\beta}}(y)
	+\Sigma^{\mu\nu}(x)K^{\Sigma\Sigma}_{\mu\nu\rho\sigma}(x-y)z^{-1}
	\frac{\delta}{\delta j^{\Sigma}_{\rho\sigma}}(y)\\
	+&c^\rho(x)K^{c\bar{c}}_{\rho\sigma}(x-y)z^{-1}
	\frac{\delta}{\delta j^{\bar{c}}_\sigma}(y)
	+\bar{c}^\rho(x)K^{\bar{c}c}_{\rho\sigma}(x-y)z^{-1}
	\frac{\delta}{\delta j^c_\sigma}(y))
\end{align}
The factors $z^{-1}$ stand for the inverse residues of the respective 
propagators. The reference to which one, we have suppressed for notational      
convenience.\\
A rather explicit construction of the Fock space for the spin two fields 
can be found in \cite{InAb}.

The Hilbert space for the $\phi^{\mu\nu}$ quartets is defined following 
\cite{KuOj}.
Amongst the states $|\phi,c,\bar{c},b>$, made up by the fields indicated, one 
selects those which are annihilated by the BRST-charge $Q\,$: $Q|\rm{phys}>=0$.
Since $\Sigma^{\mu\nu}$ is $s$-invariant its Fock space which contains negative 
norm states is still part of it. All of them build up the state 
$\mathcal{V}_{\rm phys}$.
The norms of all states not containing $\Sigma^{\mu\nu}$'s is known
to be non-negative.\\
We now recall that the $\Sigma^{\mu\nu}$ fields are projected to zero in higher 
orders. This is due to the fact that their original real poles in the tree 
approximation have been shifted on the real axis and into the complex $p^2$
plane and changed their singularity character. This change we could not prohibit 
via (symmetric) counter terms, because those are not available. They have been 
used for fixing the
symmetric invariants $\int \sqrt{-g}(c_1 R^{\mu\nu}R_{\mu\nu}+c_2 R^2)$.
In tree approximation there are, however, still nonvanishing contributions.
One might be tempted to put there ``by hand'' $c_1=c_2=0$,
with the argument that in tree approximation no higher derivatives are required. 
But this is in conflict with the solutions $g_i(t)$ of the RG-equation 
(\ref{rcpls}) which then vanish.\\
Hence one has to live with some loss of probability in tree approximation: 
All initial states made up
from $\phi^{\mu\nu}$'s which go into final states made up from $\Sigma^{\mu\nu}$'s
prohibit that positivity is realized. One can consult in this context the paper
\cite{InAb}, where (although with another aim in mind) explicitly
such processes have been studied and one can see that the higher derivatives
play already in tree approximation the important role of damping amplitudes.
E.g.\ pure EH + scalar leads to scattering amplitudes which grow too fast
for large momenta.\\
Beginning with one-loop the $\Sigma$-states can no longer be excited as
outgoing states, hence there one has as final states the above described quartet 
states.\\
It is to be noted, that in internal lines of diagrams the field $\Sigma$ is 
present and plays its growth limiting role, since there $h=\phi+\Sigma$ and the 
propagators still have their UV-fall-off with $(p^2)^{-2}$.\\
The internal lines consist of $<\phi\phi>+<\Sigma\Sigma>$ (+ other  members 
of the quartet). In the appendix of pap II we discussed how the optical theorem 
can be realized.\\

Hence the $\Sigma^{\mu\nu}$ fields and their interactions are not irrelevant as 
far as physics is concerned.
The transition amplitudes of the $\phi^{\mu\nu}$ fields amongst their quartet 
states will in general depend on the couplings $c_1,c_2$ and thereby exhibit the 
influence of the ``shadow world'' spanned by $\Sigma^{\mu\nu}$'s.

\section{Discussion and Conclusions} \label{sec:conclusions}
The present paper extends two previous one's by including a massless
scalar field as representative of matter. This is important because
gravity lives, of course, essentially with and from matter and conversely.
In particular the inclusion of higher derivative terms on the side of
gravity gains full credit from considering the interplay with matter.\\
Power counting renormalizability is valid from the outset, but nevertheless
convergence has to be and has been shown, because this is the backbone of
the BPHZL subtraction scheme which we employ. The proof shows that it can
be easily extended to cover also the presence of spin one and spin one half
fields as long as respective gauge invariance is not spontaneously broken.
The broken symmetry case requires dedicated, model dependent considerations
as presented, e.g.\ in \cite{EK98} for the electroweak standard model.\\

For the case at hand we discuss the standard machinery: invariant partial 
differential equations. First the Lowenstein-Zimmermann equation which shows 
that at the value $s=1$ all dependence from $M$, the auxiliary mass, vanishes 
for Green's functions. 
($s$ is an auxiliary subtraction variable in the scheme which is needed
to avoid off-shell infrared divergences which otherwise would be introduced
by Taylor subtractions with respect to momenta.). Second, the renormalization 
group equation is established and solved. The solution is easy in Landau gauge 
because there finiteness properties hold. As further result we derive rigid Weyl
invariance which replaces in the presence of gravity dilatations, respectively
the Callan-Symanzik equation of ordinary flat spacetime.\\

The main problem of this model is the (necessary) presence of higher
derivatives. We show that also in the presence of matter, beginning with one 
loop the massive 
poles in the propagator of the gravitational field $h^{\mu\nu}$ can not be 
reached, hence
are projected away when going on the mass shell. In tree approximation they
contribute however and cause loss of probability. This rather counterintuitive
result (one would expect troubles rather in higher orders) leads us to the 
speculation that already at the classical level some
physical effect should be at work and cure this defect. For instance,
in presence of Kerr black holes with
their very peculiar ergosphere this problem might be solved: the positive
energy field associated with massless poles would be swallowed by the black 
hole,
the negative energy field associated with the massive poles would be converted
to positive energy and scattered to spatial infinity. Of course, such a bold
speculation would require a detailed study.\\

\begin{appendix}
\section{Appendix}
\subsection{Notations and conventions}
In this work, we are employing the conventions below, which are the ``timelike 
conventions'' of Landau-Lifschitz.% (cf. \cite{Misner:1974qy}).
\[
\renewcommand{\arraystretch}{1.5}
\begin{array}{llcl}
{\rm flat \,\, metric}&\eta^{\mu\nu}&=&{\rm diag}\,\,\,(+1,-1,-1,-1)\\
{\rm Christoffel}&\Gamma^\sigma_{\mu\nu}&=&\frac{1}{2}
             g^{\sigma\rho}(\partial_\nu g_{\rho\mu}
                           +\partial_\mu g_{\rho\nu}-\partial_\rho g_{\mu\nu})\\
{\rm Riemann}    &\tensor{R}{^\lambda_{\nu\rho\sigma}} &=&\partial_\rho  \Gamma^\lambda_{\nu\sigma}
                           -\partial_\sigma\Gamma^\lambda_{\nu\rho}
                           +\Gamma^\lambda_{\tau\rho}  \Gamma^\tau_{\nu\sigma}
                           -\Gamma^\lambda_{\tau\sigma}\Gamma^\tau_{\nu\rho}\\
{\rm Ricci}&R_{\mu\nu}&=&\partial_\sigma\Gamma^\sigma_{\mu\nu} 
                        -\partial_\nu\Gamma^\sigma_{\mu\sigma}
                        +\Gamma^\sigma_{\mu\nu} \Gamma^\rho_{\sigma\rho}
                        -\Gamma^\rho_{\mu\sigma}\Gamma^\sigma_{\nu\rho}\\
{\rm curvature\,\,scalar}&R&=&g^{\mu\nu}R_{\mu\nu}                             
\end{array}
\]

\subsubsection{ Projection operators} 
In order to cope with the spin properties of the field $h^{\mu\nu}$ it is 
useful to introduce projection operators. 
They are based on the transverse and longitudinal projectors for vectors
\be\label{provec}
     \theta_{\mu\nu}\equiv \eta_{\mu\nu}-\frac{p_\mu p_\nu}{p^2} \qquad \qquad
\omega_{\mu \nu} \equiv \frac{p_\mu p_\nu}{p^2}
\ee
 the projectors are defined as 
\begin{eqnarray}\label{projs}
P^{(2)}_{{\rm TT}\mu\nu\rho\sigma} &\equiv&
 \frac{1}{2}(\theta_{\mu\rho}\theta_{\nu\sigma} +\theta_{\mu\sigma}\theta_{\nu\rho})
                                -\frac{1}{3}\theta_{\mu\nu}\theta_{\rho\sigma}\\
P^{(1)}_{{\rm SS}\mu\nu\rho\sigma} &\equiv&
	\frac{1}{2}(\theta_{\mu\rho}\omega_{\nu\sigma} 
	+\theta_{\mu\sigma}\omega_{\nu\rho} +\theta_{\nu\rho}\omega_{\mu\sigma}
	                   +\theta_{\nu\sigma}\omega_{\mu\rho})\\
P^{(0)}_{{\rm TT}\mu\nu\rho\sigma} &\equiv& 
		   \frac{1}{3}(\theta_{\mu\nu}\theta_{\rho\sigma})\\ 
P^{(0)}_{{\rm WW}\mu\nu\rho\sigma} &\equiv& 
		 \omega_{\mu\nu}\omega_{\rho\sigma}\\ 
P^{(0)}_{{\rm TW}\mu\nu\rho\sigma} &\equiv&
		 \frac{1}{\sqrt{3}}\theta_{\mu\nu}\omega_{\rho\sigma}\\ 
P^{(0)}_{{\rm WT}\mu\nu\rho\sigma} &\equiv&
		 \frac{1}{\sqrt{3}}\omega_{\mu\nu}\theta_{\rho\sigma} .
\end{eqnarray}
They satisfy the closure relation
\be\label{clsre}
(P^{(2)}_{TT} + P^{(1)}_{SS} + P^{(0)}_{TT} + P^{(0)}_{WW})_{\mu\nu\rho\sigma} =
   \frac{1}{2}(\eta_{\mu\rho}\eta_{\nu\sigma}+\eta_{\mu\sigma}\eta_{\nu\rho}).
\ee 
\end{appendix}

\subsection*{Acknowledgement}
The author is grateful to Steffen Pottel for helpful discussions in the course
of writing this paper.\\

\bibliographystyle{alpha}
\bibliography{operator_weyl}

\end{document}